\documentclass[graybox]{svmult}

\usepackage{mathptmx}       
\usepackage{helvet}         
\usepackage{courier}        
\usepackage{type1cm}        
\usepackage{makeidx}         
\usepackage{graphicx}        
\usepackage{multicol}        
\usepackage[bottom]{footmisc}
\usepackage{hyperref}        
\usepackage{soul}            
\usepackage{ulem}
\hypersetup{colorlinks=true,urlcolor=blue}

\usepackage[numbers,sort&compress,authoryear]{natbib}

\usepackage{siunitx}
\usepackage{mathtools}
\usepackage{amsmath}
\usepackage{amssymb}

\makeindex             

\newcommand{\dd}{\mbox{d}}
\newcommand{\gfrac}[2]{\displaystyle\frac{#1}{#2}}
\newcommand{\annee}{\mbox{year}}

\usepackage[usenames,dvipsnames,x11names]{xcolor}
\usepackage{tikz}
\usepackage{amsmath}
\usetikzlibrary{arrows,decorations,shapes,calc}

\definecolor{cyanscint}{rgb}{0.0,0.85,0.95}
\definecolor{redscint}{rgb}{0.95,0.3,0.0}
\definecolor{greenscint}{rgb}{0.05,0.95,0.05}

\definecolor{darkgreen_col}{rgb}{0.0,0.45,0.0}

\colorlet{scintillator}{cyanscint!60}
\colorlet{scintillator_empty}{cyanscint!15}
\colorlet{darkgreen}{darkgreen_col!95}
\colorlet{scintillator_pol}{greenscint!20}
\colorlet{scintillator_neg}{redscint!25}
\colorlet{scintillatoredge_empty}{cyanscint!95}
\colorlet{scintillatoredge}{cyanscint!99}
\colorlet{blueoptphoton}{cyan!60!blue}

\usetikzlibrary{shapes,snakes}
\usetikzlibrary{arrows.meta}
\usetikzlibrary{calc,patterns,angles,quotes}

\begin{document}
\title*{Gamma-Ray Polarimetry}
\author{Denis Bernard, Tanmoy Chattopadhyay, Fabian Kislat\thanks{corresponding author} and Nicolas Produit }
\institute{Denis Bernard \at LLR,
Ecole polytechnique, CNRS / IN2P3 and Institut Polytechnique de Paris, 91128 Palaiseau, France, \email{denis.bernard@in2p3.fr}
\and
Tanmoy Chattopadhyay \at Kavli Institute of Particle Astrophysics and Cosmology, Stanford University, 452 Lomita Mall, Stanford, CA 94305, USA \email{tanmoyc@stanford.edu}
\and
Fabian Kislat \at University of New Hampshire, 8 College Rd, Durham, NH 03824, USA \email{fabian.kislat@unh.edu}
\and
Nicolas Produit \at University of Geneva, Astronomy Department 16, Chemin Ecogia CH-1290 Versoix, Switzerland \email{Nicolas.Produit@unige.ch}}
%
%
\maketitle
\abstract{
While the scientific potential of high-energy X-ray and gamma-ray polarimetry has long been recognized, measuring the polarization of high-energy photons is challenging.
To date, there has been very few significant detections from an astrophysical source.
However, recent technological developments raise the possibility that this may change in the not-too-distant future.
Significant progress has been made in the development of Gamma-ray Burst (GRB) polarimeters and polarization sensitive Compton telescopes.
A second-generation dedicated GRB polarimeter, POLAR-2, is under development for launch in 2024, and COSI a second-generation polarization sensitive Compton Telescope has been selected by NASA for launch in 2025.
This chapter reviews basic concepts and experimental approaches of scattering polarimetry of hard X-rays to MeV $\gamma$-rays, and pair production polarimetry of higher-energy photons.
\keywords{Gamma-rays, Polarization, Instrumentation, Compton polarimetry, Pair production polarimetry, Gamma Ray Bursts, Black hole accreting systems, Neutron stars}
}

\section{Introduction}
The polarization of X-rays and $\gamma$-rays provides important observables, in addition to imaging, spectroscopy, and timing.
While this potential has been recognized for a long time, 
measuring the polarization of high-energy photons is technically challenging.
In fact, the only detection of $\gamma$-ray polarization at ${\geq}5\sigma$ significance was \textsl{AstroSat}'s measurement of the Crab \citep{vadawale17}.
However, in the last two decades there has been tremendous progress in the development of polarization-sensitive $\gamma$-ray detectors and several highly promising missions are currently under consideration and under development.

\begin{figure}[b]
    \centering
    \includegraphics[width=.5\linewidth]{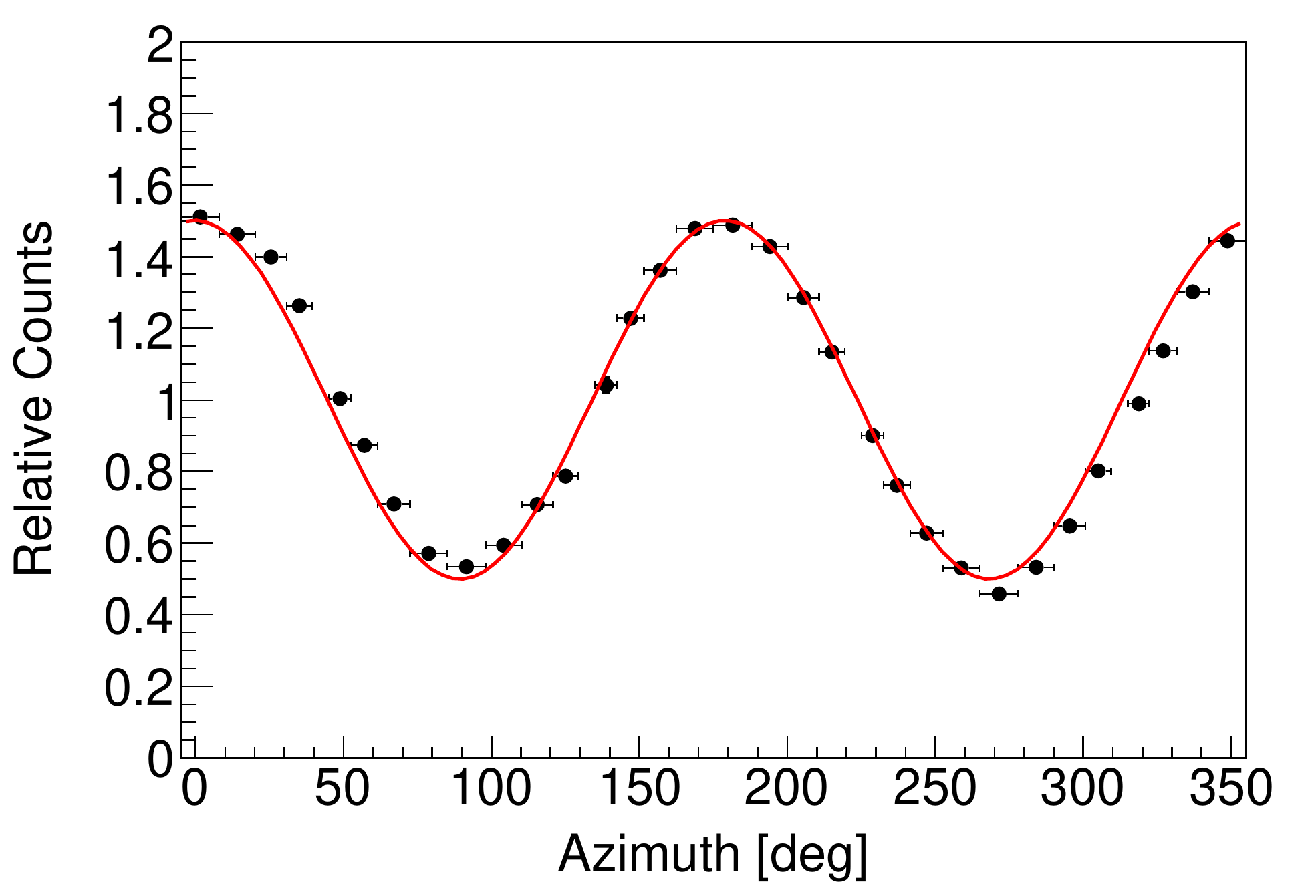}
    \caption{Azimuthal scattering angle distribution of an \SI{80}{keV} beam with a polarization fraction of \SI{85}{\percent} at the Cornell High-Energy Synchrotron Source (CHESS). With the particular event selection, the instrument has a modulation factor of about \num{0.6}. Data from~\citet{beilicke14}.}
    \label{fig:xcalibur-azimuth}
\end{figure}

Gamma-ray polarimetry covers a broad energy range from tens of keV to GeVs.
No single experimental approach can cover this entire energy range, nor is there a single approach suitable to the broad range of scientific and observational objectives.
In the energy range up to a few MeV, the azimuthal dependence of the Compton scattering cross section can be exploited, whereas at higher energies the kinematics of electron-positron pair production is used.
This is illustrated in Fig.~\ref{fig:xcalibur-azimuth} and discussed in more detail in Sections~\ref{sub:Compton:basics} and~\ref{sub:pair:xsection}.
In essence, the measured azimuthal distribution is modulated sinusoidally with a period of \ang{180}, an amplitude proportional to linear polarization fraction, and a phase related to polarization angle.
At lower energies the emission direction of photo-electrons can be used to measure polarization.
However, that is outside the scope of this chapter.
The ideal instrument design depends on whether the target of observation is a steady, predictable source, or an unpredictable transient, such as gamma-ray bursts (GRBs).
This paper gives an overview of the basic concepts of Compton and pair production polarimetry, and compares the predominant experimental approaches and their trade-offs.

While there is some astrophysical interest in circular polarization of $\gamma$-rays~\citep[e.g.,][]{Elagin:2017cgu,Huang:2019ikw,Shakeri:2018qal,Yang:2017pww,Heyl:2018kah}, there is currently no sufficiently sensitive experimental approach to measure circular polarization.
In principle, both Compton scattering~\citep{clay_hereford_1952,tashenov_2011,beard_rose_1957,trautmann_etal_1977} and pair production~\citep{OlsenMaximon,KolbenstvedtOlsen,Olsen:1959zz,Gakh:2012zz} are sensitive to circular polarization, for example when conversion or scattering takes place on polarized electrons (e.g. in magnetized iron targets), or when the polarization state of the final state electron(s) can be measured.

The complete polarization state can be described by the four Stokes parameters $I$ (intensity), $Q$ and $U$ (linear polarization), and $V$ (circular polarization)~\citep{stokes_gg_1852,kislat_etal_aph_2015b}.
The linear polarization state is fully described by $I$, $Q$ and $U$, or, alternatively, polarization fraction and angle,
\begin{equation}
  p = \frac{\sqrt{Q^2 + U^2}}{I}
  \quad\text{and}\quad
  \psi = \frac{1}{2}\arctan\left(\frac{U}{Q}\right).
\end{equation}
Due to the current lack of viable detector designs for astrophysics, this chapter focuses entirely on the measurement of linear polarization.

A key challenge of $\gamma$-ray polarimetry is that source fluxes are typically low and backgrounds high.
For example, assuming a not uncommon signal-to-background ratio of \num{1}, on the order of \numrange{1e6}{1e7} photons must be collected to detect a \SI{1}{\percent} polarized signal at the \SI{99}{\percent} confidence level.
The reason for this is that from a statistical perspective only the \SI{1}{\percent} of polarized photons contribute to the signal.
Furthermore, the relative amplitude of the azimuthal modulation may not reach \SI{100}{\percent} for a fully polarized signal.
An important characteristic of any instrument is the so called modulation factor for a \SI{100}{\percent} polarized signal,
\begin{equation}
  \mu_{100} = \frac{A_\text{max}-A_\text{min}}{A_\text{max}+A_\text{min}},
\end{equation}
where $A_\text{max}$ and $A_\text{min}$ are the maximum and minimum of the azimuthal distribution, respectively.
This modulation factor is determined by the kinematics of the photon interaction, and by the ability of the instrument to measure the inherent azimuthal modulation.
The latter depends on the process by which azimuth is measured and the geometry of the instrument.
Figure~\ref{fig:xcalibur-azimuth} illustrates the azimuthal modulation measured with the scattering polarimeter X-Calibur during a beam test at the Cornell High-Energy Synchrotron Source (CHESS).
The beam is about \SI{85}{\percent} polarized and the instrument has a modulation factor of \num{\sim 0.6}.
Typical modulation factors are in the $\mu_{100} = (\numrange[range-phrase=-]{10}{50})\,\%$ range.

The statistical sensitivity to polarization of an observation is determined by the modulation factor, signal and background event rates, and observation time.
The minimum detectable polarization (MDP) is commonly defined as the polarization fraction below which \SI{99}{\percent} observations fall in the absence of any polarization of the signal,
\begin{equation}
  \text{MDP} = \frac{4.29}{\mu_{100}R_S}\sqrt{\frac{R_S + R_B}{T}},
\end{equation}
where $R_S$ and $R_B$ are signal and background event rates, respectively, and $T$ is observation time.

Polarization is an inherently positive quantity.
While the resulting statistical challenges are well-understood, this also leads to systematic issues that must be understood in detail.
For example, any azimuthal asymmetry of the instrument can lead to a false positive signal if not mitigated or treated correctly in the data analysis.
One mitigation strategy is to rotate the instrument during observations.
However, that is not always possible.
Statistical treatment of asymmetries requires their precise knowledge, which is typically gained from detailed Monte Carlo simulations, often using the Geant4 framework \citep{Agostinelli:2002hh}.
Of course, any simulation result must be verified through laboratory measurements with both polarized and unpolarized sources.

The polarization sensitivity energy range of most Compton polarimeters lies in the sub-MeV domain, as both the Compton cross section and the Compton polarization asymmetry are decreasing functions of energy.
For pair polarimeters, for photons with an energy close to threshold 
($2 m_e c^2 \approx \SI{1}{\mega\electronvolt}$) 
the limiting factor is the ability to detect, reconstruct, select and trigger on pairs of low-momentum tracks, as the cross section decreases and the angular resolution of the telescope degrades at low energy.
Therefore, the energy range between \SI{1}{MeV} and \SI{\sim 10}{MeV} will remain the most challenging for $\gamma$-ray polarimetry.

A $\gamma$-ray polarimeter is not only a physical instrument, it involves also a simulation that is required to be able to transform the measured quantities in usable scientific measurements.
This simulation has to be checked, tuned and validated by using polarized sources in the laboratory.
The instrument also has to be fully calibrated before the launch in conditions as close as possible to those on orbit.
The ride aboard a rocket has the danger of disturbing the instrument due to depressurization, vibrations, shocks and temperature excursions.
Thus, calibration obtained on Earth must be verified and monitored in space.
Energy response and detection efficiency can in general be monitored using radioactive sources. 
However, verification of the polarization response in space is complicated by the fact that most gamma-ray emission is expected to be at least partially polarized, resulting in a lack of a reliable unpolarized standard.
Furthermore, while there are $\gamma$-ray polarization measurements of the Crab nebula, it certainly cannot be considered a true established calibration reference.
Finally, it is very hard to accurately simulate the instrumental background conditions in space.
A typical approach, thus, is to model measured backgrounds in order to understand their origin, and then incorporate those results in the data analysis.

Knowledge of the instrument response, typically obtained from simulations, is summarized in Response Matrix Files (RMFs) and Ancillary Response Files (ARFs).
These response files fully characterize the instrument and are used in the analysis of measured data.
The Geant4 simulation library \citep{Agostinelli:2002hh} accurately simulates polarized Compton scattering and can be used to derive the necessary response files, provided a sufficiently accurate mass model is implemented.
For pair conversion, after it was found~\citep{Gros:2016zst} that the existing polarized 
``Physics Model" fails to generate events with the known polarization asymmetry \citep{Gros:2016dmp}, 
a new ``G4BetheHeitler5DModel'' Physics Model \citep{Igor:2018,Bernard:2018hwf}
that samples the 5D polarized Bethe-Heitler differential cross section \citep{May1951} 
has been implemented and has been available since Geant4 release 10.6 \citep{Ivanchenko:2020wvu}.

This paper is structured as follows.
Section~\ref{sec:science} provides an overview of the scientific motivation for $\gamma$-ray polarimetry.
Basic concepts and experimental approaches to scattering polarimetry are discussed in Section~\ref{sec:Compton}, and pair-production polarimetry is reviewed in Section~\ref{sec:pair}.
The paper closes in Section~\ref{sec:summary} with a summary, discussion of unresolved issues, and future outlook.

\section{Science drivers of Gamma ray polarimetry}\label{sec:science}
Science drivers for polarimetry in energies ranging from a few tens of keV to the MeV regions, focuses a number of X-ray sources that are bright in this energy range. Scientific potential in hard X-rays and the recent findings from some of the recent X-ray spectro-polarimetric instruments have been summarized in \citet{chattopadhyay21_review}. The following is a brief description of the science cases for some of the bright gamma ray sources.

\begin{itemize}
 \item Gamma Ray Bursts: Measurement of Gamma ray polarization has the potential to test the proposed models for GRB prompt burst emission mechanism  \citep{covino16,mcconnell16,toma08}, e.g., synchrotron 
emission from relativistic electrons either in a globally ordered magnetic field \citep{lyutikov03,nakar03,granot03} or in a random magnetic 
field generated in the
shock plane within the jet \citep{medvedev99} or emission due to Comptonization
of soft photons \citep{shaviv95} by the relativistic jet. In the past decade, several dedicated X-ray polarimetry experiments and some of the spectroscopic instruments have reported measurements of X-ray/$\gamma$-ray polarization for  several GRBs (see reviews by \citet{mcconnell16,gill21_grbpol,chattopadhyay21_review}). Gamma-Ray Burst Polarimeter (GAP) and CZT Imager (CZTI) aboard \textsl{AstroSat} reported high polarization fractions ($>$50 \%) for a few GRBs \citep{yonetoku11,yonetoku12,chattopadhyay19}. For CZTI, GRB intervals were optimized for best detection of polarization. On the other hand, POLAR, a dedicated GRB polarimeter aboard Chinese space station, found most of the GRBs to have low or null polarization in their full burst intervals \citep{zhang19,Kole20polar_catalog}. Although a firm conclusion on the prompt emission requires polarization of a larger number of GRB sample, the existing results indicate that GRB prompt emission is highly structured which can lead to possible variation in polarization angle within a burst as has been seen in a number of GRBs \citep{vidushi19,Burgess_etal_2019}.     
 \item Active Galactic Nuclei (AGN): The power-law component seen in AGN spectra corresponding to the coronal emission is expected to be polarized owing to scattering of disk photons by corona. Measurement of polarization of this component can be useful in investigating the corona geometry \citep{schnittman10}. 
 \item Blazars:  In case of low energy peaked blazars, polarization in MeVs is expected to explain the origin of the high energy peak in their spectral energy distribution. For example, while the Synchrotron Self Compton model \citep[SSC model,][]{celotti94} predicts around 30 \% polarization in a uniform magnetic field, the External Compton 
model (EC) predicts null polarization ($<$5 \%) for this component \citep{mcnamara09}. 

Gamma-ray polarization has the potential to also distinguish between the leptonic and the lepto-hadronic models of $\gamma$-ray emission by blazars.
In the hadronic models the high energy peak is believed to originate from the high energy proton induced synchrotron radiation.
So while in the (2 -- 10 keV) X-ray band, the maximum (i.e. assuming a magnetic field without turbulence) polarization fraction is predicted to be similar for the two families of models, in the (30 -- 200 MeV) $\gamma$-ray band,
 the hadronic models predict a much higher level of polarization fraction than the leptonic models~\citep{zhang13}.

 \item Accreting stellar mass black holes:
 In the hard spectral state of the accreting black hole systems, the coronal emission (dominant in $\sim$10-100 keV) is expected to be polarized owing to scattering of thermal disk photons by the corona with polarization characteristics significantly influenced by the coronal geometry \citep{schnittman10}. 
 At energies beyond 100 keV, polarization measurements might be useful to investigate models proposing jet origin of high energy radiation and unravel the 
various aspects of jet emission in black hole sources like disk-jet connection, jet
launching mechanism, jet composition (see \citep{shahbaz16_v404,Kantzas20}). Lately, there are some interesting reports of polarization measurements in hard X-rays for the famous high mass black hole X-ray binary system, Cygnus X-1. \textsl{PoGO+}, a balloon borne polarimeter, measured low polarization (with an upper limit in polarization fraction of 5.6 \%) in 20-180 keV in its hard state observed between 12 $-$ 18 July, 2016  \citep{chauvin18a,chauvin18b}, suggesting the X-ray photons below 200 keV to originate from an extended corona. Imager on Board the INTEGRAL Satellite (IBIS) and Spectrometer on INTEGRAL (SPI) on board \textsl{INTEGRAL}, on the other hand, independently estimated  high polarization ($>$50 \%) for Cygnus X-1 at energies above $\sim$300 keV, while the emission at relatively lower energies was found to be weakly polarized \citep{laurent11,jourdain12}. 

 \item Accreting neutron stars:
 Phase resolved polarization study for the accreting neutron stars in hard X-rays can be useful in determining the beam shape of the pulsar e.g., to distinguish between the pencil and
fan beam radiation patterns where oscillations in polarization fraction with the pulse phase are predicted to be opposite to each other \citep{meszaros88}. In case of millisecond X-ray pulsars, polarization measurement has the potential to test the origin of high energy photons and put tighter constraints
on geometrical parameters like orbital and dipole axis inclinations \citep{viironen04,sazonov01}. 
Recently, the balloon-borne hard X-ray polarimeter \mbox{\textsl{X-Calibur}}, during a balloon flight in December 2018, reported a phase-integrated \SIrange{15}{35}{keV} polarization fraction of $27_{-27}^{+38}\,\%$ for  GX 301-2 as well as on-pulse and bridge polarizations of $32_{-32}^{+41}\,\%$ and $27_{-27}^{+55}\,\%$, respectively, which does not constitute a non-zero detection~\citep{abarr20}.

 \item Magnetars:
 Gamma ray polarization for magnetars can test the RICS model \citep[Resonant Inverse Compton Scattering, see][]{wadiasingh17,Beloborodov12,baring06} explaining the origin of hard X-ray tail seen in the spectra of a several magnetars. A high level of linear polarization in hard X-raysbe a direct confirmation of RICS process in magnetar's lower magnetosphere \citep{wadiasingh19_white}.
 Polarimetry studies of magnetars also offer unique opportunity to test the QED effects in strong magnetic field.
 \item Pulsars:
 Phase-resolved $\gamma$-ray polarization has the potential to investigate the emission sites of high-energy radiation from rotation powered 
pulsars. The theoretical models e.g., polar cap model \citep{daugherty82}, outer gap model \citep{cheng00}, slot gap model \citep{dyks03}, and striped pulsar wind \citep{kirk02,petri05} propose different acceleration sites for the relativistic electrons and predict distinct phase dependent polarization signatures which can be tested by gamma ray polarimetry studies of X-ray pulsars. In recent times, there have been a number of reports on measurement of polarization of Crab pulsar and nebula in hard X-rays from instruments like SPI, IBIS, PoGO+, CZTI, Soft Gamma ray Detector (\textsl{SGD}) on board \textsl{Hitomi} and \textsl{POLAR}~\citep{Jourdain19,forot08,chauvin18,vadawale17,hitomi18,Hancheng2021}. All these instruments found high polarization ${\gtrsim}\SI{20}{\percent}$ for both phase averaged Crab and nebula separately with polarization angle closely aligned with the pulsar spin axis, 124.0$\pm$0.1$^\circ$~\citep{ng04}. 

 \item Solar flares:
 Gamma-ray polarization measurement of solar flares has the potential to investigate the emission mechanism behind the high-energy photons which is widely believed to be from non-thermal Bremsstrahlung emission by the high energy electrons. Polarization measurements are expected to probe multiple crucial model parameters like beaming of the electrons, magnetic field structure, back-scattering of the photons from the 
photosphere and dependence of polarization on the location of the flare on the disk \citep{bai78,leach83,zharkova10,jeffrey11}. Reuven Ramaty High Energy Solar Spectroscopic Imager (\textsl{RHESSI}) and a polarimeter \textsl{SPR-N} on board \textsl{CORONAS-F} reported polarization for a sample of M and X class flares in hard X-rays~\citep{garcia06,zhitnik06}. Due to the large uncertainties in the measurements, any firm conclusion on the hard X-ray origin of the flares was not feasible from these measurements.
\end{itemize}

\section{Scattering Polarimetry}\label{sec:Compton}

\begin{table}
\caption{
  Hard X-ray and $\gamma$-ray polarimetry missions, important prototypes, and proposed missions and key characteristics, representing a broad cross-section of past and current experimental efforts. The order is approximately chronological. The numbers in the table are only indicative, the energy range and the effective area may be different for polarization measurements and in case of balloons may not include absorption in atmosphere. ``Type'' lists the type of mission and of the instrument according to the classification used in Section~\ref{sec:scattering:approaches}.
  }
\label{tab:scattering:missions}
\scriptsize
\begin{tabular}{ ||c|c|c|c|c|c|c|c|c||}
 \hline
 Instrument & FOV & Energy & Effective & Technology & Angular    & Type & Science & Status \\
            &     & range  & Area      &            & Resolution &&&\\
 \hline\hline
 SPR-N&Full Sun disk& \SIrange{20}{100}{keV} & \SI{50}{\centi\meter^2} &Be scatterer&Non imaging&Satellite&Solar flares&2001-2005\\
 &&&& and scintillators&&&&\\
 \hline
 MEGA &\SI{2\pi}{\steradian}& \SIrange{300}{5e4}{keV} &\SI{324}{\centi\meter^2}& Silicon strips &Some degree& Balloon &GRB& Prototype \\
 &&&& and CsI && 3D &Point sources&\\
 \hline
 PHENEX&$\ang{4.8}$& \SIrange{40}{200}{keV} & \SI{44}{\centi\meter^2} &Plastic scintillator&Non imaging&Balloon&Point source&2006 and 2009\\
 &&&& and CsI&&Collimated&&\\
 \hline
 TIGRE&$\ang{45}\times\ang{45}$& \SIrange{400}{1e5}{keV} &\SI{80}{\centi\meter^2} at\SI{1}{MeV}& Silicon strips & \ang{2} ARM & Balloon &GRB& 2010 \\
 &&&& and CsI & at \SI{1}{MeV} & 3D &Point sources&\\
 \hline
 PENGUIN-M&$\ang{45}\times\ang{45}$& \SIrange{20}{150}{keV} & \SI{78}{\centi\meter^2} & Plastic scintillator & Non imaging&Satellite&Solar flares&2009-2010\\
   &&&& and NaI&&&&\\
 \hline
 GAP&$\ang{30}\times\ang{30}$& \SIrange{50}{300}{keV} & \SI{176}{\centi\meter^2} &Plastic scintillator&Non imaging&Satellite&GRB&2010-2011\\
  &&&& and CsI&&Wide-field&&\\
 \hline
 GRAPE& \SI{2\pi}{\steradian}& \SIrange{50}{500}{keV} & \SI{144}{\centi\meter^2} &Plastic scintillator&Non imaging&Balloon&GRB&2011 and 2014\\
   &&&& and CsI && Wide-field &&\\
 \hline
 POGO+&$\ang{2}$& \SIrange{20}{180}{keV} & \SI{1400}{\centi\meter^2} &Plastic scintillator&$\ang{2}$&Balloon&Point sources&2016\\
   &&&& and CsI && Collimated &&\\
 \hline
 COSI &\SI{\pi}{\steradian}& \SIrange{200}{2000}{keV} & \SI{256}{\centi\meter^2} &Segmented Ge&\ang{3.2}&Balloon& GRB & 2016\\
 (Balloon) &&&&&& 3D & Point sources & \\
 \hline
 SGD&$\ang{0.6}\times\ang{0.6}$ & \SIrange{50}{200}{keV} & \SI{210}{\centi\meter^2} &Si pixels & \ang{30} & Satellite & Point sources&2016\\
 &&&& and CdTe&& Collimated &&\\
 \hline
 X-Calibur /& \ang{;6;} & \SIrange{20}{60}{keV} & \SI{10}{cm^2} at \SI{50}{keV} &Be scatterer&Non imaging&Balloon&Point sources&2016, 2019, \\
 XL-Calibur && \SIrange{20}{80}{keV} & \SI{100}{cm^2} at \SI{50}{keV} & and CZT&& Focal plane && 2022 \\
 \hline
 INTEGRAL & $\ang{9}\times\ang{9}$ & \SIrange{15}{1e4}{keV} & \SI{2600}{\centi\meter^2}  & CdTe, CsI & \ang{;;12} &Satellite&Point sources& Flying\\
 IBIS &&&&&& Coded mask &GRB& since Oct 2002 \\
 \hline
 INTEGRAL & $\ang{9}\times\ang{9}$ & \SIrange{15}{1e4}{keV} & \SI{500}{\centi\meter^2}  & Ge & \ang{1} &Satellite&Point source&Flying\\
 SPI &&&&&& Coded mask && since Oct 2002 \\
 \hline
 AstroSat & $\ang{4.6}\times\ang{4.6}$ & \SIrange{100}{350}{keV} & \SI{924}{\centi\meter^2}  & CZT &Non imaging &Satellite& GRB, & Flying\\
 &&&above \SI{100}{keV} &&&Coded mask& Point sources &since 2015 \\
 \hline
 POLAR & \SI{2\pi}{\steradian} & \SIrange{50}{500}{keV} & \SI{300}{\centi\meter^2}   & Plastic scintillator & \ang{10} & Space station & GRB & 2016-2107\\
 & & &at \SI{300}{keV} && bright GRB & Wide-field &&\\
 \hline
 POLAR-2 & \SI{2\pi}{\steradian} & \SIrange{10}{500}{keV} & \SI{1250}{\centi\meter^2} &Plastic scintillator & \ang{5} & Space station & GRB & Manifested\\
 &&&at \SI{300}{keV} &&bright GRB& Wide-field &&2024\\
 \hline
 PING-P &$\ang{45}\times\ang{45}$ & \SIrange{20}{150}{keV} & \SI{30}{\centi\meter^2}  &Plastic scintillator & Non imaging &Satellite&Solar flares&2025\\
 & & & &and CsI &&&&\\
 \hline
 PolariS &$\ang{;;10}\times\ang{;;10}$& \SIrange{10}{80}{keV} &\SI{3.2}{\centi\meter^2} &Plastic scintillator &\ang{1}  & Satellite & Point sources & Under development \\
 & & & &and GSO && Focal plane && Launch TBD \\
 \hline
 COSI &\SI{\pi}{\steradian} & \SIrange{200}{2000}{keV} & \SI{256}{\centi\meter^2} &Segmented Ge &\ang{3.2} & Satellite & GRB, &Selected\\
 (SMEX) & & & & & & 3D & Galactic sources & 2025 \\
 \hline
 LEAP & \SI{1.5\pi}{\steradian}&\SIrange{50}{500}{keV} & \SI{1000}{\centi\meter^2}&Plastic scintillator & $1 -- \ang{5}$ & ISS & GRB & Proposed \\
  & & & &and CsI && Wide-field &&\\
 \hline
 AMEGO &$\SI{2.5}{\steradian}$& \SIrange{200}{1e6}{keV} &$\SI{608}{\centi\meter^2}$ &Silicon strips &\ang{2.5}& Satellite &Point sources& Proposed \\
 & & & &CZT and CsI &at $\SI{1}{MeV}$& 3D &GRB&\\
 \hline
 e-ASTROGAM &$\SI{2.5}{\steradian}$& \SIrange{300}{3e6}{keV} &$\SI{10000}{\centi\meter^2}$ &Silicon strips & \ang{0.15} & Satellite &Point sources& Proposed \\
 ASTROMEV & & & &and CsI & at $\SI{1}{GeV}$& 3D &GRB&\\
 \hline
\end{tabular}

\textbf{References:} SPR-N \citep{bogomolov03}, MEGA~\citep{bloser_2006}, PHENEX \citep{gunji08}, TIGRE  \citep{oneill_1996,bhattacharya_2004}, PENGUIN-M \citep{dergachev09}, GAP \citep{yonetoku06}, GRAPE \citep{bloser09}, POGO+ \citep{friis18}, COSI (Balloon) \citep{yang18}, SGD \citep{hitomi18}, X-Calibur \citep{kislat18}, XL-Calibur \citep{abarr21}, IBIS \citep{ubertini03}, SPI \citep{vedrenne03}, AstroSat \citep{vadawale15}, POLAR \citep{produit18}, POLAR-2 \citep{kole19}, PING-P \citep{kotov16}, PolariS \citep{hayashida14}, COSI (SMEX) \citep{tomsick19}, LEAP \citep{mcconnell21_leap}, AMEGO \citep{amego19}, e-ASTROGAM \citep{de18}
\end{table}

\subsection{Basic Concepts}\label{sub:Compton:basics}
Rayleigh and Compton scattering, like essentially all electromagnetic interactions, conserve polarization.
Because photons are massless, their polarization is perpendicular to their momentum vector.
As an immediate consequence, photons preferentially scatter perpendicular to their polarization.
In fact, in the low-energy limit a \ang{90} scattered photon scattering along the polarization direction would lead to a longitudinally polarized photon, which is forbidden.
Rayleigh scattering and photo effect dominate at low energy and are more important for material of high atomic number $Z$. Compton scattering dominates for energies between \SI{\sim 50}{keV} and \SI{\sim 10}{MeV}, depending on the target material.

The cross section of the Compton process (see figure \ref{compton}) follows the Klein-Nishina formula~\citep[e.g.][]{evans_r_1955}:
\begin{equation}\label{KN}
\frac{d\sigma}{d\Omega}=\frac{1}{2} r_{e}^{2}\left(\frac{E^{\prime}}{E}\right)^{2}\left[\frac{E}{E^{\prime}}+\frac{E^{\prime}}{E}-2 \sin ^{2}(\theta) \cos ^{2}(\phi)\right],
\end{equation}
with
\begin{equation}\label{KNE}
1+\left(E / m_{e} c^{2}\right)(1-\cos \theta)=\frac{E}{E^\prime}=\frac{\lambda^{\prime}}{\lambda}
\end{equation}
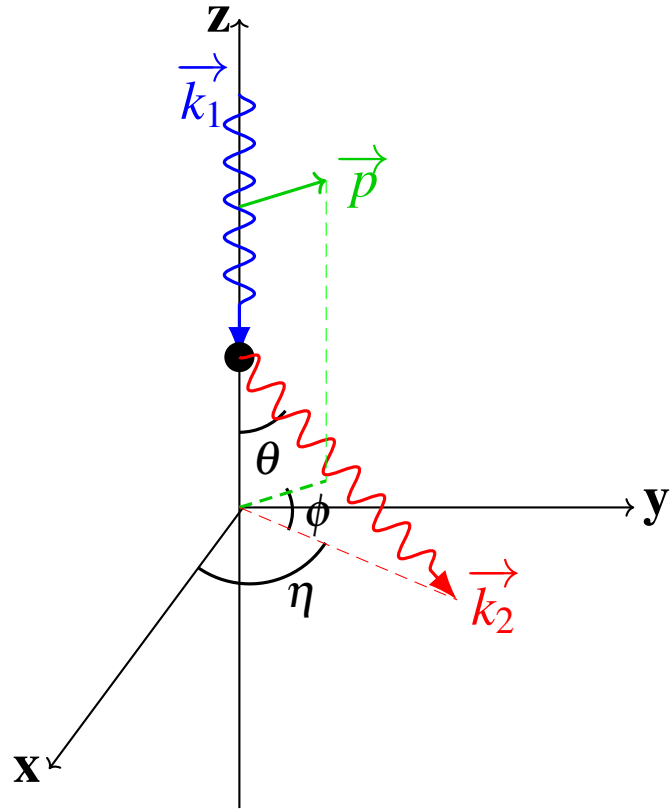
\begin{figure}[b]
\centering
\begin{tikzpicture}
\pgfmathsetmacro{\cubex}{0.5}
\pgfmathsetmacro{\cubey}{6}
\pgfmathsetmacro{\cubez}{0.5}
\pgfmathsetmacro{\offsetx}{2}
\pgfmathsetmacro{\offsety}{-2}
\pgfmathsetmacro{\offsetz}{-2}

\draw [black, thick, <-] (-\cubex/2,3.5,-\cubez/2) node[left]{{\huge\textbf z}} -- (-\cubex/2,-7,-\cubez/2);
\draw [black, thick, ->] (-0.2,-3,-\cubez/2) -- (5,-3,-\cubez/2) node[right]{{\huge\textbf y}};
\draw [black, thick, ->] (-0.2,-3,-\cubez/2)-- (-2.3,-6,1) node[left]{{\huge\textbf x}};
\coordinate (origin) at (-\cubex/2,-3,-\cubex/2);

\coordinate (scat1) at (-\cubex/2,-1,-\cubez/2);
\coordinate (scat2) at(\offsetx-\cubex/4,\offsety-3,\offsetz-\cubez/2);

\draw [blue, very thick,-{Latex[length=4mm]},decorate, decoration={snake,amplitude=2mm,segment length=5mm,post length=5mm}, label=left:Incident photon] (-\cubex/2,2.5,-\cubez/2) node[left]{{\huge $\overrightarrow{\boldsymbol{k_1}}$}} -- (scat1);

\fill[black] (scat1) circle (0.2);

\draw [black, very thick] (-\cubex/2,-2,-\cubez/2) node[black, below]{{\LARGE\hspace*{0.6cm} $\theta$}} arc (270:320:0.8) ;

\draw [black, very thick] (0.4,-3.3,-\cubex/2) arc (-25:30:0.6) ;
\draw (0.5,-3.1,-\cubex/2) node[black, right]{{\LARGE $\phi$}};

\draw [black, very thick] (-0.8,-3.8,-\cubez/2) arc (235:327:1.2) ;
\draw (0.2,-3.9,-\cubez/2) node[black, below]{{\LARGE\hspace*{0.6cm} $\eta$}};

\draw [red, very thick,-{Latex[length=4mm]},decorate, decoration={snake,amplitude=2mm,segment length=5mm,post length=5mm}, label=left:Scattered photon] (-\cubex/2,-1,-\cubez/2) -- (scat2) node[right]{{\huge $\overrightarrow{\boldsymbol{k_2}}$}};

\draw [green!80!black, very thick,->] (-\cubex/2,1,-\cubez/2) -- (1-\cubex/2,1.2,-\cubez/2-0.4) node[right]{{\huge $\overrightarrow{\boldsymbol{p}}$}};

\draw [green, thin, dash pattern=on5pt off3pt] (1-\cubex/2,1.2,-\cubez/2-0.4) -- (1-\cubex/2,-2.8,-\cubez/2-0.4);
\draw [green!80!black, very thick, dash pattern=on5pt off3pt] (-\cubex/2,-3,-\cubez/2) -- (1-\cubex/2,-2.8,-\cubez/2-0.4);
\draw [red, thin, dash pattern=on5pt off3pt] (origin) -- (scat2);

\end{tikzpicture}

\caption{Definition of the different axis and angles, a Compton scattering takes place at the black dot, a photoelectric effect takes place at the end of the $k_2$ vector}
\label{compton}
\end{figure}%
\begin{figure}[t]
\centering
\includegraphics[height=.375\linewidth]{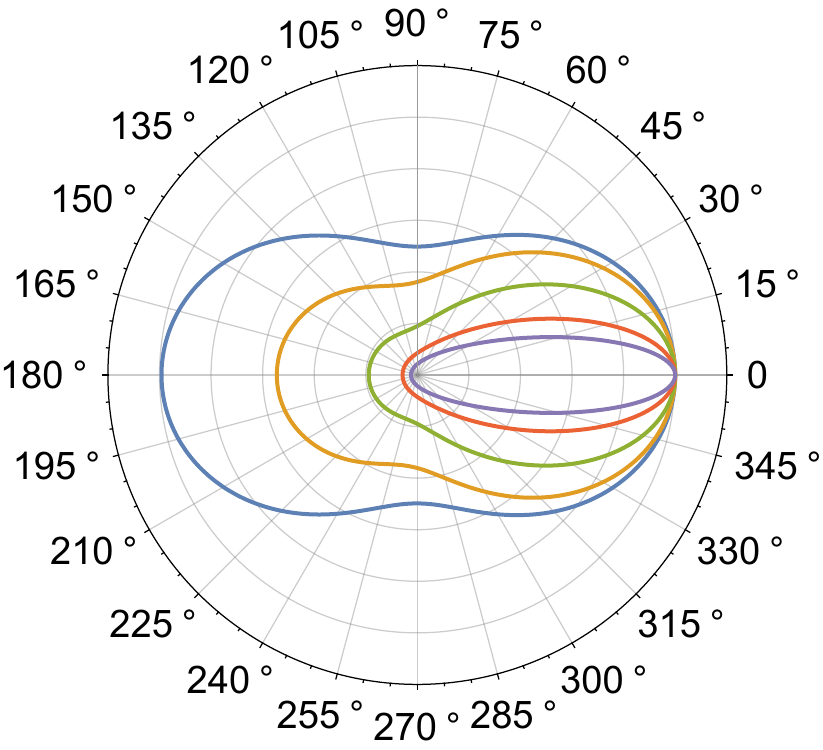}%
\hfill%
\includegraphics[height=.375\linewidth]{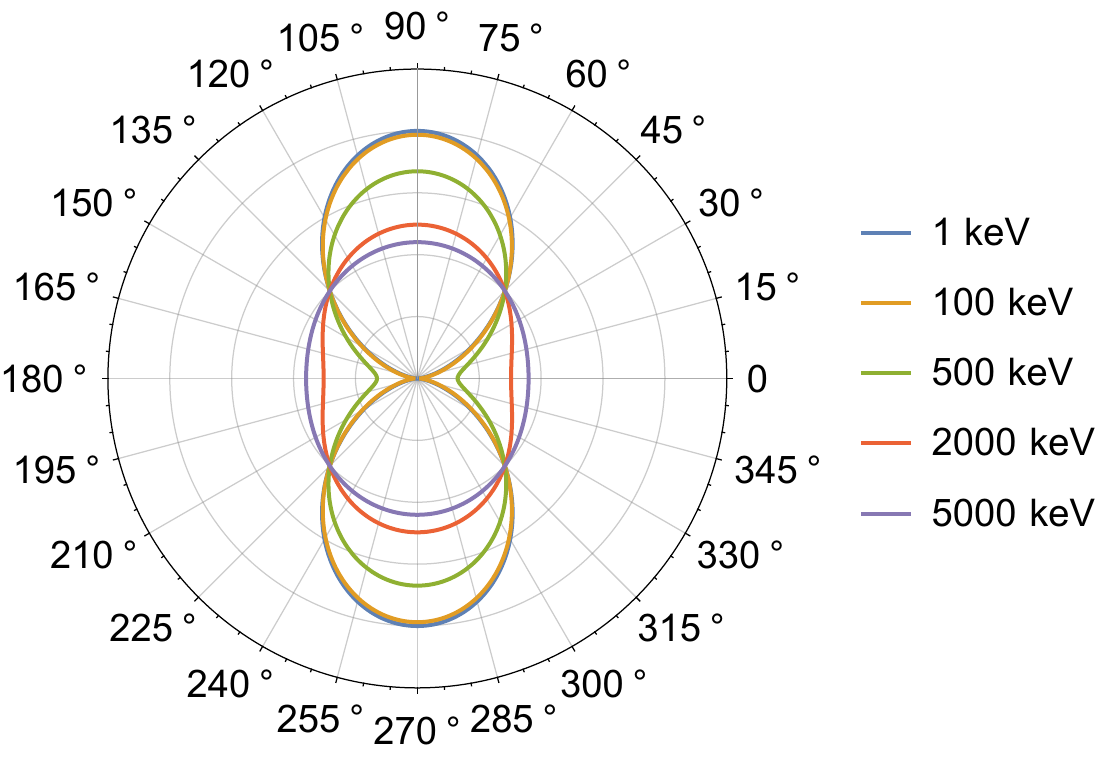}%
\caption{%
  Illustration of the Klein-Nishina cross section~\eqref{KN}.
  \emph{Left:} Scattering angle $\theta$ distribution for photons incident from the left.
  \emph{Right:} Azimuthal scattering angle $\phi$ distribution for photons incident normal to the page scattering by~$\theta = \ang{90}$.
}
\label{fig:klein-nishina}
\end{figure}%
Here, $r_e = e^2/4\pi\epsilon_0m_ec^2 \approx \SI{2.818e-15}{\meter}$ is the classical electron radius, E, $\lambda$, $E'$, $\lambda'$ are the energy and wavelength before and after scattering, respectively, $\theta$ is scattering angle, $\phi$ is azimuthal scattering angle with respect to polarization
(See  Fig.~\ref{compton}).
The cross section is illustrated in Fig.~\ref{fig:klein-nishina}.
It depends on polarization through the $\phi$ angle, which must be measured by any scattering polarimeter.
The dependence is in $\cos^2\phi$ (or $2 \phi$ as $\cos^2 \phi= (\cos 2\phi+1)/2$) as it should be because polarization lives in the projective plane so must have a $\pi$ symmetry.
The cross section is strongly forward-backward enhanced by the $\sin^2\theta$ term, with a strong forward-enhancement in the high-energy limit.
The polarization sensitivity is strongest when $\theta$ is $\pi/2$.
Hence, there is an advantage to have a detector that can detect perpendicular scatterings.
The direction of the recoil electron in Compton scattering is also sensitive to polarization.
While this is being exploited by some experimental approaches (see Sec.~\ref{sub:scattering:approaches:3d}), directional information of the electron is rapidly lost due to multiple scattering.

\subsection{Experimental approaches}\label{sec:scattering:approaches}
\begin{figure}[t]
  \centering
  \includegraphics[width=.6\linewidth]{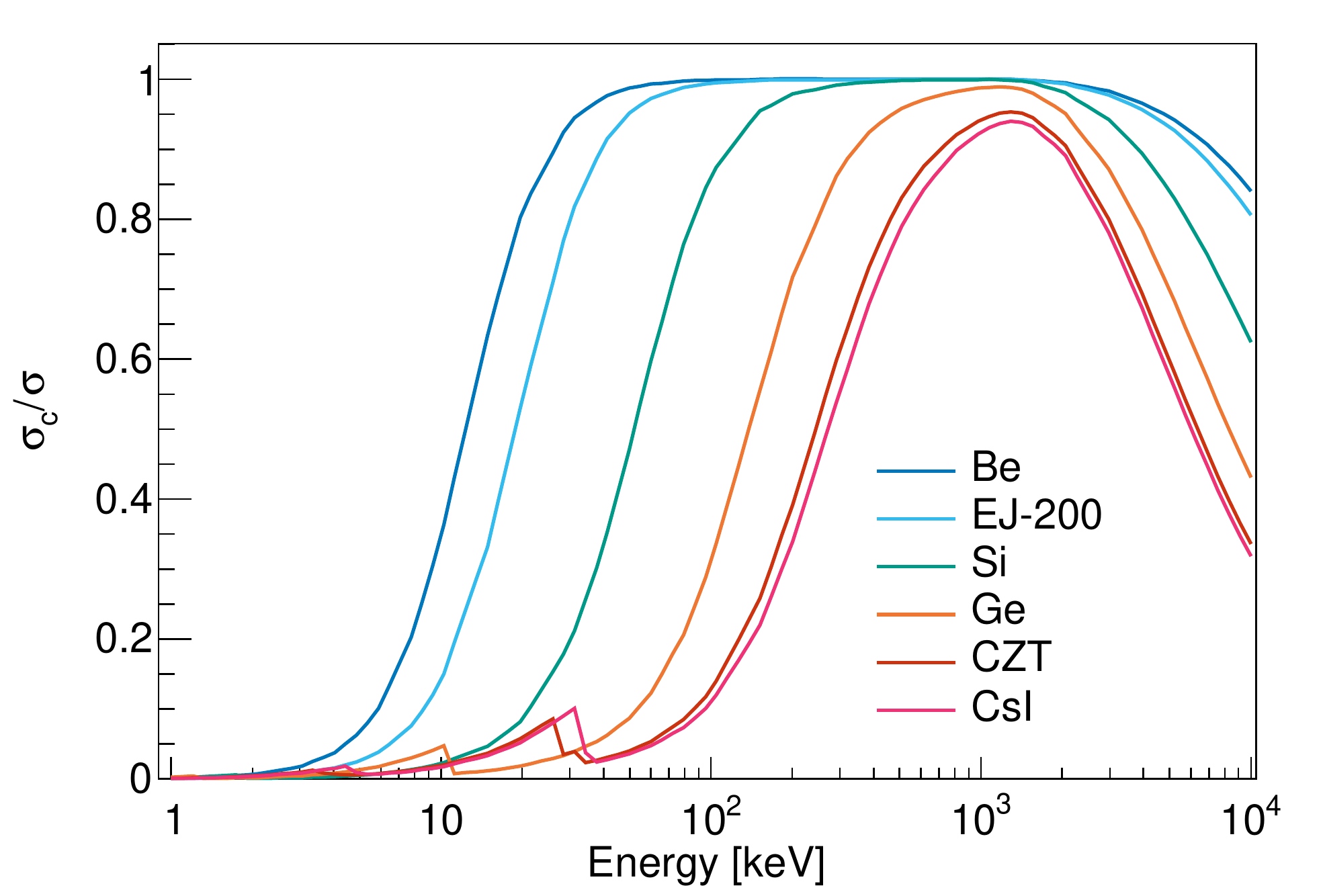}
  \caption{Ratio of scattering to total cross section of a selection of commonly used materials. EJ-200 (base material polyvinyltoluene) is shown as an example of a common plastic scintillator material. At low energy photoelectric absorption dominates, whereas above a few MeV electron-positron pair production dominates.
  Data from the NIST XCOM photon cross-section database.}
  \label{fig:scattering-cross-section}
\end{figure}

A range of instruments based on these general principles have been designed, built, tested and in some cases flown on spacecraft or high-altitude balloons (Table~\ref{tab:scattering:missions}).
Differences in instrument design result from specific scientific objectives, as well as a range of detailed trade-offs.
One of the most important considerations in the design of a Compton polarimeter is the selection of detector materials.
This is illustrated in Fig.~\ref{fig:scattering-cross-section}, which shows the fractional cross section for scattering for a selection of commonly used materials.
Low-Z materials are ideally suited as scattering target over a broad energy range.
In fact, below \SI{\sim 20}{keV} purity of the scattering material is critically important and even a small contamination with high-Z elements can degrade performance.
On the other hand, high-$Z$ materials such as inorganic CsI scintillator or CZT semiconductor detectors provide a large photon absorption cross section though it should be noted that scattering is the dominant interaction process over an energy range from a few hundred keV to a few MeV, regardless of target material.
As a consequence, many detector designs consist of dedicated low-Z scattering and high-Z detector elements, although some instruments use only a single material.

Compton polarimeters can broadly be divided into two classes: wide field and point source instruments.
Wide field instruments are ideally suited for studies of bright transient events such as gamma-ray bursts, and for extended sources.
Wide-field instruments can further be subdivided into wide field polarimeters and Compton telescopes, which in addition to polarimetry have significant imaging capabilities.
Point source instruments are optimized for in-depth observations of point-like sources.
Their background tends to be lower resulting in a higher sensitivity for steady sources.
Point source instruments can further be subdivided into collimated large area instruments and focal plane instruments.
The latter utilize imaging or light collecting optics allowing a significant reduction of the detector size, and as a consequence background, while maintaining a significant effective area.
While the optics tend to have less effective area than large area polarimeters, the improved signal-to-background ratio enhances sensitivity to weak sources and objects with low polarization fraction.
The following sections describe examples of each of these four classes, and the trade-offs that led to the various designs.
The instruments were chosen based on their relevance to the field and to highlight certain aspects and trade-offs of each implementation.
This section is not intended to represent an exhaustive list.

\subsubsection{Wide field instruments}
Wide-field $\gamma$-ray polarimeters are characterized by a field of view that covers a significant fraction of the sky, typically on the order of $(1-2)\SI{\pi}{sr}$, and a large photon collection area.
This class of $\gamma$-ray polarimeters is primarily designed to measure the polarization of short transient events, such as gamma-ray bursts and magnetar outbursts.
The unpredictable nature and short duration of these events makes the wide field of view necessary.
Another key requirement is a large effective area in order to detect a sufficient number of photons during a short event, typically a few seconds for a short GRB and up to a few hundred seconds for a long GRB~\citep{von-kienlin_etal_2020}.
The instruments discussed in this section are designed with these primary requirements in mind.
Imaging and spectroscopy are not driving factors.
All instruments discussed here are optimized for an energy range around the GRB peak flux, typically ranging from tens to hundreds of keV.

\begin{figure}[t]
\centering
\includegraphics[width=.7\textwidth]{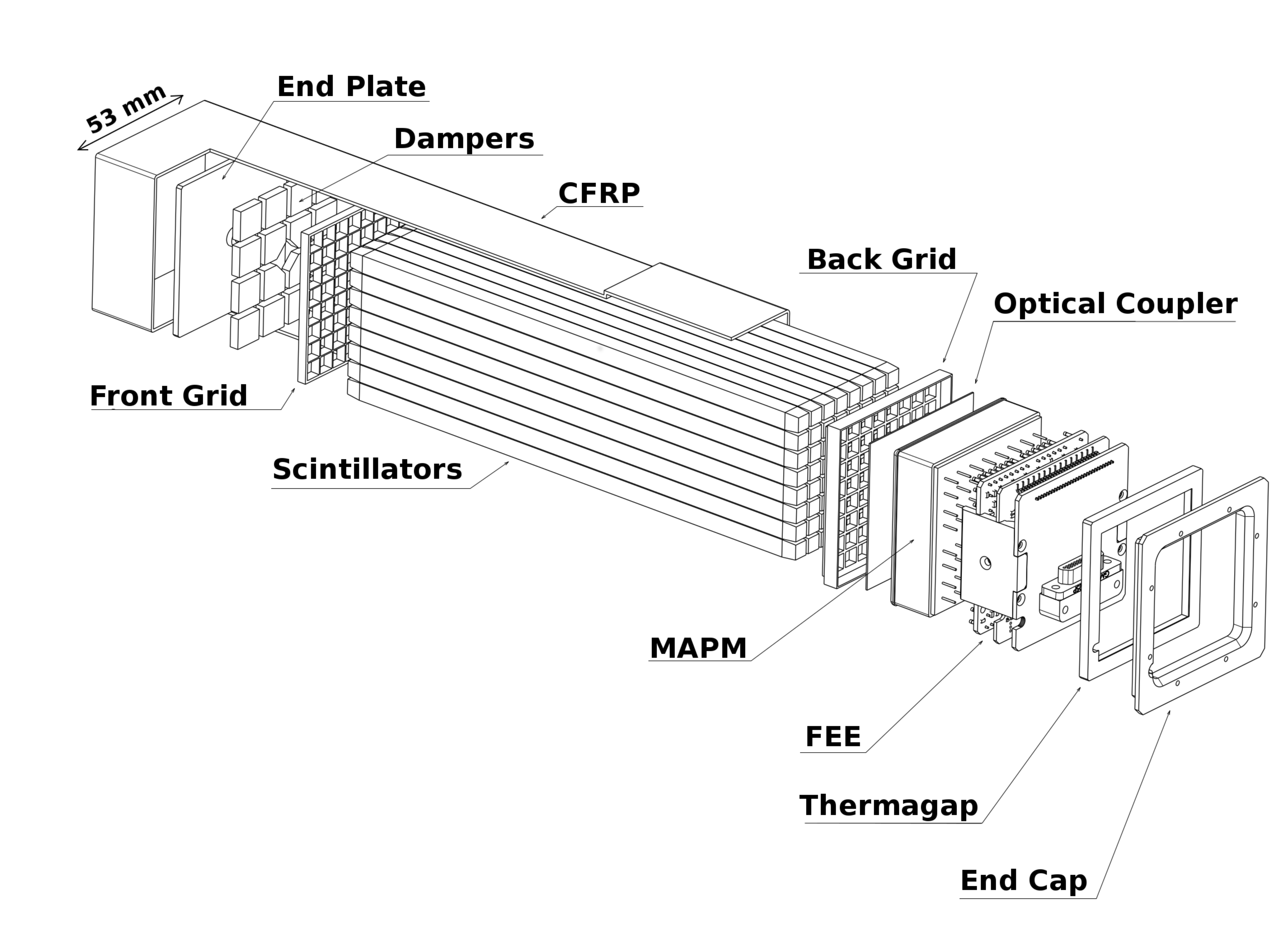}
\caption{One of the 25 modules of POLAR. The 64 plastic scintillator bars are read out by a multi-channel PMT. Triggering and analog-to-digital conversion are handled by the Front-End Electronics (FEE) board.}
\label{fig:polar_module}
\end{figure}

Several instruments in this category have been flown or are currently under development: the Gamma-Ray Burst Polarimeter \textsl{GAP}~\citep{yonetoku04} aboard the Japanese \textsl{IKAROS} satellite; \textsl{POLAR}~\citep{produit18} on the Chinese space laboratory \textsl{Tiangong2}; the balloon-borne Gamma-ray Polarimeter Experiment \textsl{GRAPE}~\citep{bloser09}; \textsl{POLAR-2}~\citep{DeAngelis2021v1} currently under construction; and the proposed Large Area Burst Polarimeter \textsl{LEAP} for the International Space Station~\citep{mcconnell21_leap}.
They are all based on an array of scintillator detectors that serve both as scattering target for the incoming photons, and as photoabsorber material for the scattered photons.
The scintillators are typically in the form of long bars read out by photomultiplier tubes or SiPMs on one end.
Figure~\ref{fig:polar_module} shows a detector module of \textsl{POLAR} consisting of 64 plastic scintillator bars, illustrating the concept.

Events in which a photon Compton scatters in one scintillator element and is absorbed in another are ideally suited to measure polarization.
In that case, determination of polarization information from the azimuthal scattering angle is completely geometrical.
Due to the $\pi$ symmetry of polarization, it is not necessary to determine the order of interactions.
On the other hand, if the incident photon is immediately absorbed in one detector element, no polarization information can be obtained.
The exact detector geometry is the result of an optimization and different conclusions may be reached depending on different requirements and priorities.

For example, \textsl{POLAR} consists of an array of 1600 plastic scintillator bars of \SI[product-units=power]{6x6x200}{mm} arranged on a square grid and read out by multi-anode photomultiplier tubes~\citep{produit18}.
The goal of this design is to maximize the fraction of photons that interact in two separate elements regardless of incident angle, creating a field of view of about~\SI{2\pi}{\steradian}.
The designers also found that this approach maximizes the mean free path between photon interactions, which improves the quality of the azimuth angle determination. Usage of very fast plastic scintillator also enable to reduce the coincidence window to \SI{50}{ns} so there are virtually no random coincidences even during most bright GRB or heavy background conditions.
The successor mission of \textsl{POLAR}, \textsl{POLAR-2}, will use the same design, but four times the number of detector modules.
Use of SiPMs instead of photomultiplier tubes (PMTs) will improve quantum efficiency and thus reduce the detector threshold to about~\SI{10}{keV}.

\textsl{LEAP}, on the other hand, consists of 7 modules of 144 scintillator bars, each, of which 84 are plastic scintillators and 60 are inorganic CsI(Tl) scintillators.
The bars are \SI[product-units=power]{17x17x100}{mm} in size, optically isolated, and read out by individual photomultiplier tubes.
The CsI elements are arranged around the edge of each module as well as forming an X in the middle.
This design limits the field of view to about~\SI{1.5\pi}{\steradian} because most photons at large zenith angles would have to pass through a CsI element before scattering in a plastic element.
On the other hand, the likelihood that scattered photons are absorbed is increased, and CsI provides a better energy resolution than plastic scintillator.

\textsl{GAP} has a better defined scattering geometry in featuring a unique plastic scintillator in the middle and a ring of high-$Z$ scintillators. This geometry enable very accurate azimuthal angle measurement but the effective area and the angular acceptance are small.

Neither design allows direct single-photon imaging capabilities.
However, all instruments have the ability to localize transient events by comparing event rates in different detector elements.
For example, the statistical localization error for a burst with an MDP \SI{<30}{\percent} in \textsl{LEAP} is in the $1-5^\circ$ range, depending on zenith angle and \textsl{POLAR} has demonstrated this technique using flight data.
This type of localization analysis requires extensive and highly detailed Monte Carlo simulations of the instrument and typically tabulation of expected count rates in each detector element, or in groups of detector elements, as a function of direction to the burst.

The same, however, is true for the interpretation of polarization data.
While it is straightforward to determine the projected azimuthal scattering angle of individual photons, the response of these types of instruments is highly asymmetrical in azimuth.
It depends not only on the direction to the burst relative to the orientation of the instrument, but may also depend on the polarization angle.
Asymmetries are determined by the precise arrangement and shape of detector elements and the presence of passive materials in the assembly.
Monte Carlo simulations with a detailed and accurate mass model are needed in order to measure this response.

\subsubsection{3D instruments}\label{sub:scattering:approaches:3d}
\begin{figure}[t]
\begin{center}
 \includegraphics[width=0.34\textwidth]{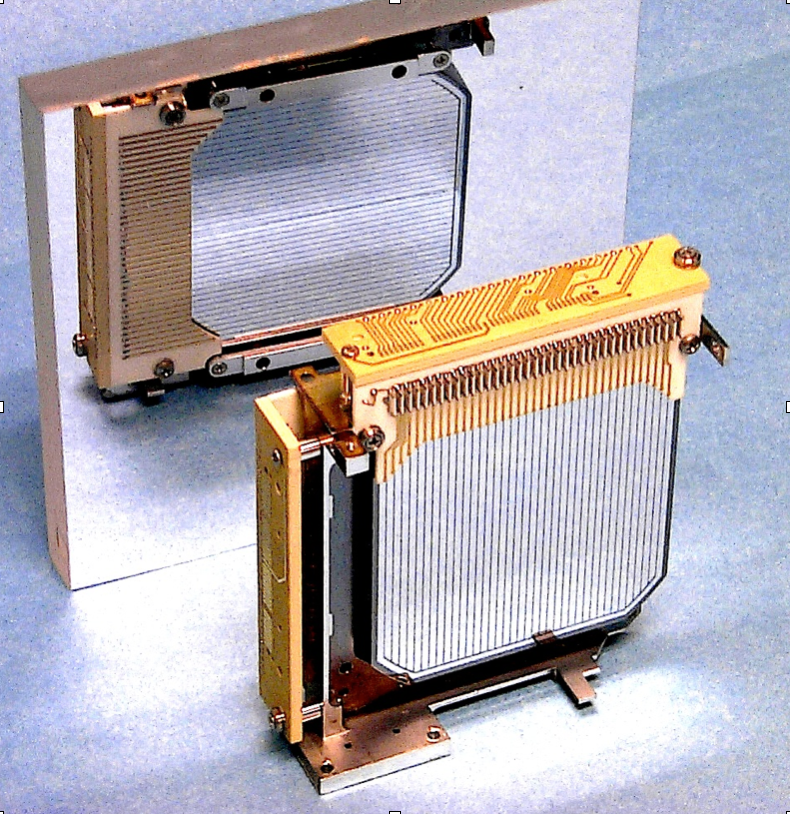}
 \hfill
 \includegraphics[width=0.65\textwidth, trim=40 20 140 0, clip]{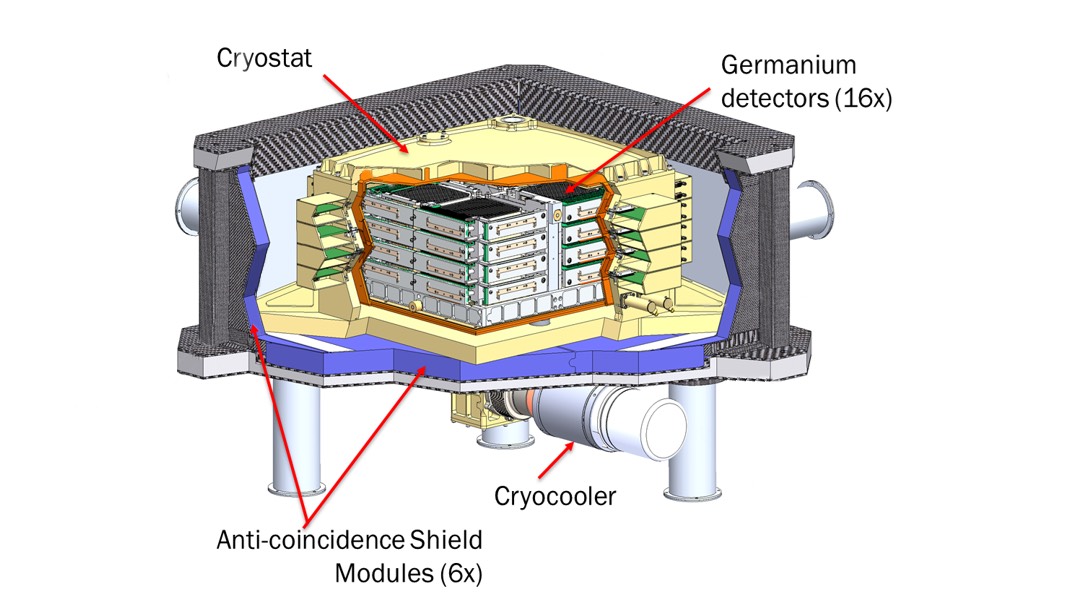}
\caption{A 3D instrument: \textsl{COSI}, the Compton Spectrometer and Imager.
Left: one of the Germanium detectors for the balloon flight,  wire bonded in it's cryostat mount;
the mirror reflection shows the orthogonal electrodes \citep[from][with permission.]{sleator_etal_2019}.
Right:
the  cryostat shown with a cutaway view to expose the 16  Germanium detectors within,   surrounded on five sides by a Bismuth Germanate (BGO) scintillator  anti-coincidence shield \citep[adapted from][]{tomsick19}. 
\label{fig:COSI}}
\end{center}
\end{figure}
Compton telescopes make use of the kinematics of Compton scattering in order to reconstruct the arrival direction of photons.
Photons scatter in one part of the detector and are photoabsorbed in another part.
By measuring both interaction locations and energy deposits, it is possible to reconstruct the origin of the photon within a circle in the sky.
If additionally the trajectory of the recoil electron emitted during Compton scattering can be measured, this circle reduces to an arc.
Point sources are then identified as locations in the sky where circles or arcs from multiple photons intersect.
Direction resolution in Compton telescopes is defined as the angular resolution measure (ARM), which is the full width half maximum of the distribution of angular separation between the Compton cone or arc and the true photon origin.
One of the biggest challenges in the analysis of data from Compton telescopes is reconstructing the correct sequence of interactions, in particular when incorporating events with more than two interaction locations.
This detection principle is inherently sensitive to polarization.
However, first-generation Compton telescopes such as \textsl{COMPTEL} consisted of two separate detectors with large separation \citep{1993Schoenfelder}.
This separation restricts the telescope to detecting photons with mostly small Compton angles, reducing polarization sensitivity, except for highly off-axis sources.

Over the course of the last two decades, new instruments with compact geometries have been developed.
These 3D detectors are sensitive to large scattering angles, and therefore, promise much greater polarization sensitivity.
A key requirement for the design of a telescope of this kind is that photon interaction locations can be determined accurately in three dimensions, as well as good energy resolution.
A particular advantage of 3D instruments over other types of polarimeters is the fact that they measure not only the azimuthal but also the polar scattering angle.
This information can be combined with the known kinematics of Compton scattering in the data analysis, and has been demonstrated to achieve a significant improvement in polarization sensitivity~\citep{krawczynski_2011,lowell_etal_2017b}.

One example of 3D instrument is the Compton Spectrometer and Imager \textsl{COSI}, which consists of a stack of three $2 \times 2$ layers of cooled Germanium detectors
(Fig. \ref{fig:COSI}).
When a photon interacts in one of the detectors, either via Compton scattering or photoabsorption, the fast recoil electron deposits its energy via ionization of the material, creating electron-hole pairs in the process.
Electrons and holes drift towards anode and cathode, respectively, in the applied electric field.
The electrodes are segmented into \SI{2}{mm} wide orthogonal strips, allowing positioning in two dimensions.
The depth of the interaction location within the detector is determined by measuring the collection time difference of signals between anode and cathode, which has been shown to be fairly linear with depth~\citep{coburn_etal_2003,amman_etal_2000}.

\begin{figure}[t]
\begin{center}
 \includegraphics[width=0.56\textwidth, trim=4 40 40 4, clip]{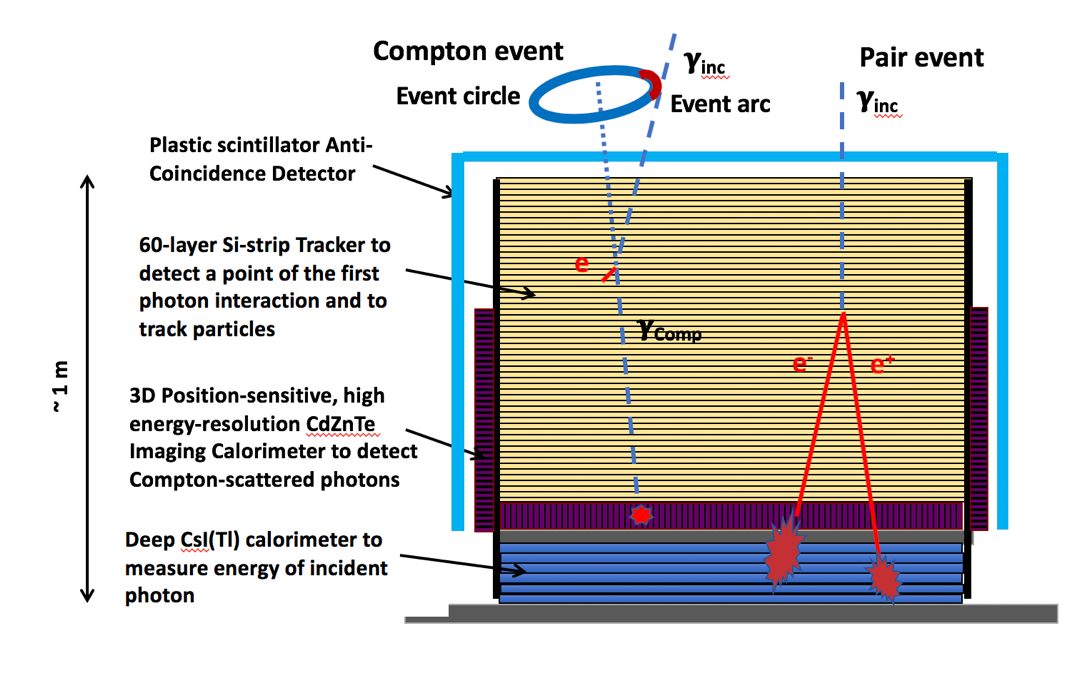}
 \hfill
 \includegraphics[width=0.42\textwidth, trim=4 4 340 4, clip]{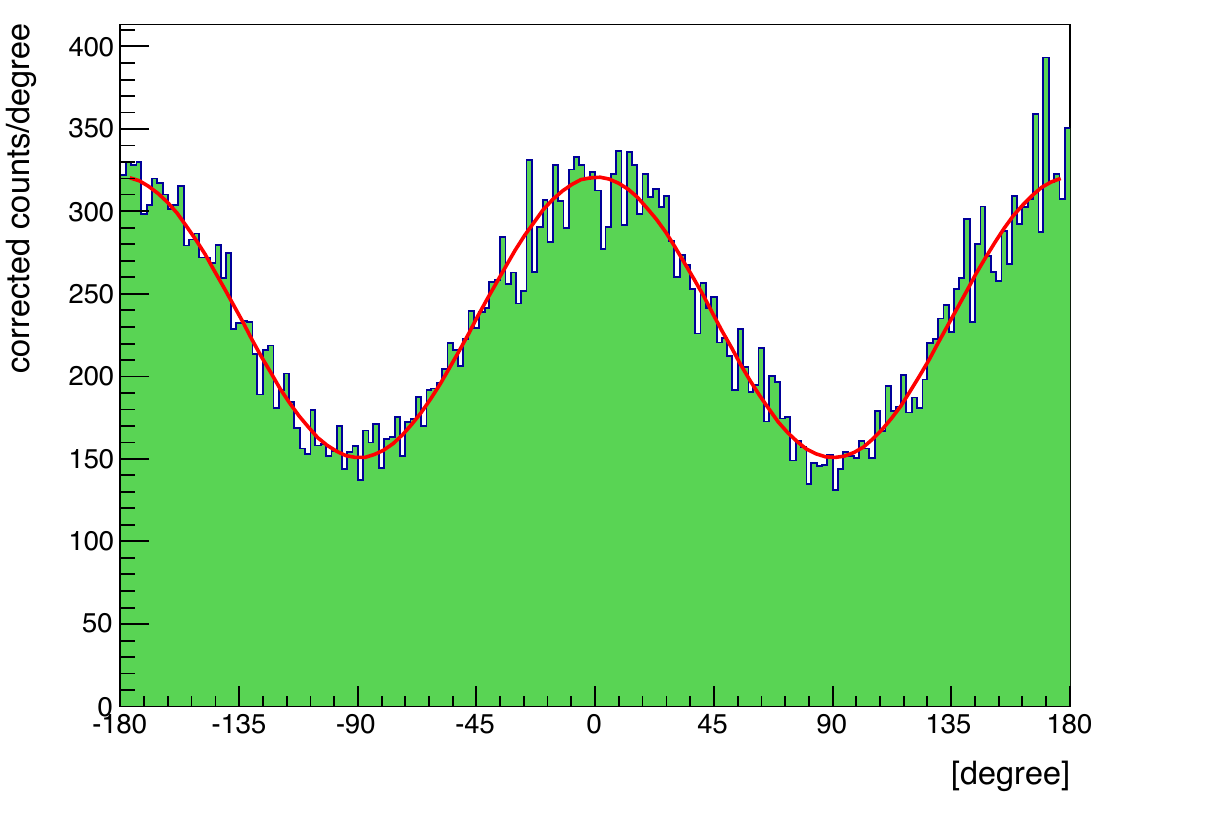}
 \caption{SSD-stack-based polarimeters.
Left:
AMEGO concept, with schematically shown Compton and pair-production
events \citep[Open Access, licence at
\href{https://creativecommons.org/licenses/by-nc-nd/4.0/legalcode}{url},][]{Moiseev:2020maz}.
Right:
e-ASTROGAM azimuthal angle distribution for Compton scattered events in the 0.2 -- 2 MeV
range for a 100\% polarized, 10 mCrab-like source observed on axis for
10$^6$ s.
The corresponding modulation is $\mu_{100}$ = 0.36 \citep[with permission,][]{de18}.
\label{fig:3D:silicon}}
\end{center}
\end{figure}

Another way of reconstructing a Compton event in 3D is by using planes of silicon detectors (Fig.~\ref{fig:3D:silicon}).
Silicon strip detectors (SSD) with double sided perpendicular readout is providing energy and 3D position of the recoil electron at each photon interaction point.
The double-sided readout is instrumental in enabling the determination of the two transverse coordinates of the position of scattered electrons that have such a low energy that they cannot exit the interaction wafer.
Those detector use anticoincidence to limit the background from charged cosmic rays and calorimeter to measure the energy of the scattered photon.

When using sufficiently thin Si detectors, the Compton recoil electron may escape the material and can be tracked through several layers.
In addition to the use of thin Si detectors, minimization of dead material between the layers is crucial to reduce scattering of the electron.
This concept was first implemented in the \textsl{MEGA} prototype \citep[\SI{500}{\micro\meter} Si,][]{bloser_2006} and the \textsl{TIGRE} balloon mission \citep[\SI{300}{\micro\meter},][]{bhattacharya_2004}.
The proposed \textsl{eASTROGAM}~\citep{DeAngelis:2016slk} and \textsl{AMEGO}~\citep{amego19} follow the same ideas
and but with a larger number of tracker planes (56 for \textsl{eASTROGAM}, 60 for \textsl{AMEGO}). 
This type of detector also enables imaging and spectroscopy at higher energies by tracking the lepton pair from pair production events.
However, it is unclear if pair polarimetry can be achieved with this approach.

While semiconductor detectors are the dominant technology used for Compton telescopes today, other approaches are also being pursued.
For example, the proposed \textsl{PANGU} concept uses scintillating fibers in place of Silicon strip detector~\citep{2016SPIE.9905E..6EW}. 
The APT mission concept uses sheets of CsI read out via wavelength shifting fibers to enable localization combined with a scintillating fiber tracker~\citep{APT:2021lhj}.
In the extreme case of a low-density active target such as a gas time-projection chamber (TPC), the tracking of the electron can be so precise that the Compton arc becomes quite short, and the scatter plane deviation becomes of the same order of magnitude as the angular resolution measure~\citep[the electron tracking Compton camera (ETCC) concept,][]{Komura:2017npl}.

\subsubsection{Collimated and coded-mask instruments}
Collimated $\gamma$-ray polarimeters are instruments with narrow field of view, typically around a few degrees. The idea is to conduct simultaneous spectro-polarimetry study (with or without coarse imaging) of bright persistent X-ray sources. Historically, Crab pulsar and nebula and the high mass black hole X-ray binary, Cygnus X-1, are the two most popular sources that have been observed a several times because of their brightness in hard X-rays/$\gamma$-rays.  

Apart from the basic need of large collecting area like in any other non-focal $\gamma$-ray polarimeters, one requirement in designing a point source polarimeter is the use of collimators to not only achieve the desired field of view and avoid source confusion but also to minimize the cosmic background in the detector. Design of the collimators is critical to block the high energy protons and neutrons or the secondary particles and photons generated during interactions of those high energy particles with the collimator and the payload support structure or the spacecraft from entering into the detector plane.
Some instruments employ graded shield in the collimators to reduce the effect of secondaries and fluorescence photons. Anti-coincidence shielding is also used in some instruments to reject the background particles.
The aperture with maximal acceptance of the collimator must be large enough to allow for inaccuracies in spacecraft pointing.
The instruments can also be employed with a coded mask on top of the collimators to enable coded mask imaging and measurement of simultaneous background which is particularly useful for spectroscopy measurements.
This technique is particularly relevant for spatially segmented detector planes, e.g., pixelated CZT detectors. However, there has not been any demonstration of measurement of simultaneous polarimetric background which can be a strong tool in the future for polarization measurements of fainter sources.
Use of coded mask only for spectroscopy can be disadvantageous for polarimetry as the mask blocks a fraction of the detector collecting area and thereby reducing the overall polarimetric sensitivity of the instrument.   

The main advantage of collimated instruments over the use of focusing optics (see next section) is that a large effective area can be maintained at high energy.
The upper energy limit of the state-of-the-art hard X-ray mirrors is limited to less than \SI{100}{keV} which makes the point source instruments to be the only realistic approach for polarimetry study of X-ray sources above \SI{100}{keV}.

In the last two decades, several dedicated point source polarimetry instruments have been launched, e.g., \textsl{PoGO+} and \textsl{PoGO} (Polarized Gamma-ray Observer) balloon borne mission, \textsl{PHENEX} (Polarimetry for High ENErgy X rays), a balloon borne mission. There are also some instruments that are primarily meant for spectroscopy but sensitive to polarimetry measurements, e.g., Soft Gamma-ray Detector (SGD) on board Japanese astronomy mission, \textsl{Hitomi}, the Imager on Board the INTEGRAL Satellite (IBIS), the SPectrometer on INTEGRAL (SPI), Cadmium Zinc Telluride Imager (CZTI) on board \textsl{AstroSat}. All these instruments widely vary in Compton scattering geometry as well as types of detectors used. Below we briefly discuss these instruments and the different geometrical configurations.
It is to be noted that some of these instruments were not calibrated for polarimetry before launch, e.g., IBIS and SPI aboard \textsl{INTEGRAL}. 
The instrument response to polarized and unpolarized radiation solely relies on simulation results. The relevant measurements, therefore, have been questioned sometimes and not considered conclusive. 

A single module of \textsl{PHENEX} \citep{kishimoto07,gunji08}, a balloon-borne point source X-ray polarimeter employs a \num{6x6} array of square shaped  plastic scintillators (\SI{5.5x5.5}{mm} in cross-section and \SI{40}{mm} in length)  
surrounded by 28 CsI (Tl) scintillators of same dimensions on four sides. The events resulting from scattering of an incoming photon by a plastic scintillator and subsequent photo-absorption by one of the surrounding CsI scintillators, are used to extract polarization information.  
Use of \SI{1}{mm} thick Molybdenum collimators restrict the FOV to \ang{4.8} (FWHM) in an energy range of \SIrange{20}{200}{keV} to perform
pointed observations. The sides are covered by graded passive shields made of \SI{2}{mm} thick Pb and \SI{1}{mm} thick Sn. The scintillators are read by a single MAPMT
(multi-anode PMT).
Crosstalk between the MAPMT pixels can reduce the polarimetric sensitivity of the instruments. While use of Si Photo-multipliers solves the cross-talk problem in the modern polarimeter designs, it can help in lowering down the energy threshold in the scintillators and thereby provide significant improvement in the sensitivity.

\begin{figure}[t]
\centering
\includegraphics[scale=.6]{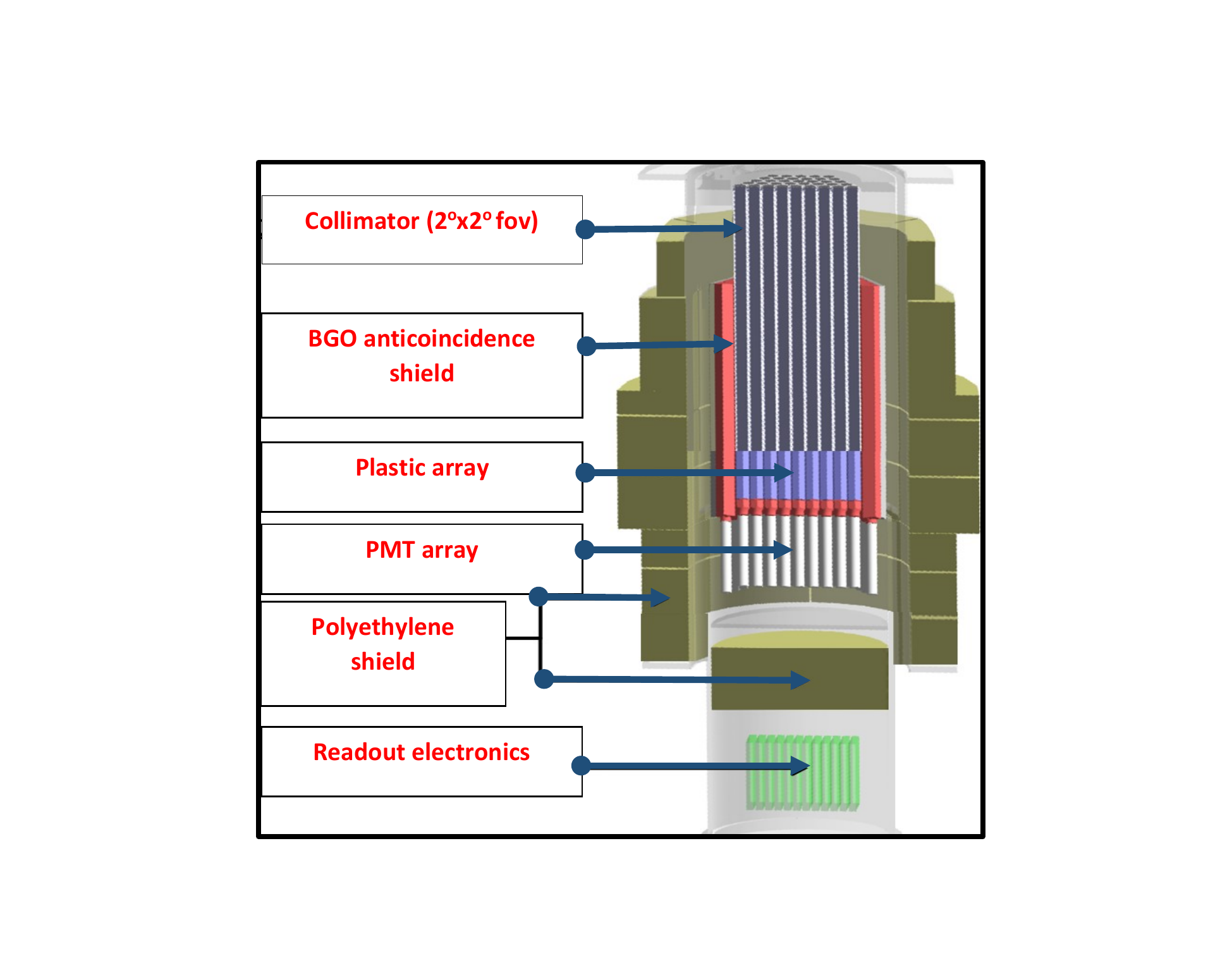}
\caption
{Schematic of the \textsl{PoGO+} instrument. It uses an array of plastic scintillators for both scatterers and absorbers. Scintillators are readout by individual photomultiplier tubes (PMTs). Use of collimators restrict the field of view of the instrument to $\ang{2}\times\ang{2}$.}
\label{pogo}
\end{figure}

On the other hand, \textsl{PoGO+}~\citep{chauvin16_pogo,friis18} utilizes an array of 61 low-$Z$ EJ-204 plastic scintillators. Each hexagonally shaped scintillator measures \SI{12}{cm} in length and \SI{3}{cm} across.
The scintillators are read by individual PMTs. 
Because this type of configuration uses only an array of low-$Z$ plastics, the plastics play a dual role of both scatterer and absorber. The advantage of this type of design is that the instrument can be scaled up easily in collecting area.
\textsl{PoGO+} was flown in 2016 in balloon borne mission for pointing observations of Crab and Cygnus X-1.
The design of \textsl{PoGO+} (shown in Fig. \ref{pogo}) is a matured version of the 
pathfinder balloon-borne experiment, \textsl{PoGOLite}, flown in 2013 \citep{chauvin16_pogolite}. 
The graded shield collimators (graded shield of Copper, Tin and Lead layers) restrict the FOV to ${\sim}\ang{2}$ for pointed observations in the \SIrange{20}{180}{keV} range. While the collimators suppress most of the off-axis background events from the front side, active BGO shielding on the sides and on the back of the scintillators vetoes most of the background events from other directions.
The estimated polarimetric sensitivity (MDP) of \textsl{PoGO+} is around \SI{11}{\percent} for Crab like sources for a 5-day balloon flight.

CZTI~\citep{bhalerao16,rao16}, aboard \textsl{AstroSat}~\citep{agrawal06,singh14} and SPI \citep{vedrenne03} aboard \textsl{INTEGRAL} offer polarimetric sensitivity utilizing their spatially segmented detector planes. A spatially segmented/pixelated spectrometer should intrinsically provide Compton polarimetry capability if the readout preserves multi-pixel or multi-detector events.
For example, CZTI, a coded mask narrow field of view instrument, employs an array of 64 pixelated CZT detector modules (each module is \SI{40x40x5}{mm} in size) with a pixel pitch of~\SI{2.5}{mm}.
Because of significant Compton scattering probability in \SI{5}{mm} thick CZT detectors above \SI{100}{keV}, CZTI offers polarimetric capability in the energy range of \SIrange{100}{380}{keV} where a valid Compton scattering event is obtained from the adjacent CZTI pixels within a coincidence window of \SI{20}{\micro\second}~\citep{chattopadhyay14,vadawale15}. 
SPI is also a coded mask instrument with 19 hexagonal high-purity Germanium detector bars (\SI{7}{cm} in height) closely packed in a hexagonal shape. Events depositing energies in two detectors within a coincidence window of \SI{350}{ns} are used to extract polarization information in the \SIrange{200}{1000}{keV} energy range~\citep{chauvin13}. An advantage of Germanium detectors is that they can be made thicker, up to a few cm, which makes them sensitive in a broad energy range with high Compton scattering efficiency. However, unlike CZT detectors, Germanium detectors require cooling to very low temperatures to minimize the leakage current.
Besides the bright persistent X-ray sources e.g., Crab and Cygnus X-1, both the instruments are also sensitive to GRB spectro-polarimetry studies, because of increasing transparency of the collimators, coded mask and the payload support structure at higher energies.

Correct interpretation of the polarization data from segmented detectors needs careful consideration of some  important factors. For example, in CZTI, because the detectors are developed on a single crystal CZT, identification of the scattering and photo-absorption event in a valid Compton scattering event can be complicated. Although, at lower energies (below \SI{260}{keV}), the electron recoil energy is always lower than the scattered photon energy, at energies beyond \SI{260}{keV}, the energy deposited in electron recoil can be larger than the scattered photon energy for certain ranges in scattering angle. On the other hand, the square pixels (typically available in pixelated detectors) introduce an asymmetry in the scattering geometry. This yields an inherent modulation in the azimuthal angle distribution even for completely unpolarized radiation.
Modulation factors, because of the asymmetry in scattering geometry, may also depend on the polarization angle \citep{chattopadhyay14}. Charge sharing between the pixels also results in fake 2-pixel events and dilutes the polarization signal.      
All these effects should be carefully evaluated for correct interpretation of data from pixelated detector based polarimetry instruments. 
Recent progress in the development of 3D position sensitive CZT detectors~\citep{zhang12_3DCZT,kuvvetli14_3dCZT}, promises a significant background reduction by constraining the source direction for each 2-pixel event. An array of 3D CZTs can therefore be used as a sensitive Compton polarimeter both in collimated and wide field configuration~\citep{caroli18}.

IBIS \citep{ubertini03} on board \textsl{INTEGRAL}, on the other hand uses a Compton camera configuration consisting of two layers of pixelated sensitive detectors.
The first layer uses CZT and the second layer uses CsI. Coincidence events between the two layers enable to reconstruct Compton events and the direction of the incoming photon on a cone. The geometry of IBIS is not optimized for polarimetry studies. The geometry only allows the forward or back scattering events with most of the 90-degree scattering events being rejected by the trigger logic. IBIS also suffers from high noise in the CsI layer and the slowness of the detector, requiring a rather large coincidence time window. 
A similar Compton camera concept is used by the SGD~\citep{tajima10} on board Hitomi~\citep{hitomi18}, but with significantly better geometrical configuration. It consists of 32 layers of \SI{0.6}{mm} thick Silicon sensors and 8 layers of \SI{0.75}{mm} thick CdTe detectors. Each detector is pixelated with a pixel size of \SI{3.2}{mm}. Two more layers of the same CdTe detectors surround these detector layers from all sides to avail the \ang{90} scattering events. Eight such units totalling to around \SI{210}{cm^2} of collecting area are collimated separately to a narrow FOV. Excellent energy and position resolution and optimized scattering geometry of SGD allow sensitive polarimetry measurements in the energy range of \SIrange{50}{200}{keV} with MDP level of a few percent for 100~mCrab sources in a \SI{1e5}{s} exposure.

\subsubsection{Focal plane instruments}
Focal plane polarimeters combine a polarization-sensitive X-ray or $\gamma$-ray detector with focusing X-ray optics. 
The advantage of this design is that the effective area is determined by the X-ray optics, not by detector size.
This allows compact instruments, which helps reduce the background count rate, and makes thick, high-$Z$ anti-coincidence detectors surrounding the instrument feasible.
It has been shown that grazing incidence X-ray optics have a \SI{<1}{\percent} impact on the polarization of reflected photons, due to the small reflection angles.
The main disadvantage is the limitation of the energy range imposed by available X-ray optics.
As such, focal plane scattering polarimeters were first proposed for the \SI{<10}{keV} X-ray energy range~\citep{kii87}.
Furthermore, the complexity of the resulting systems is much higher than for the point source instruments discussed in the previous section.

The first flown focal-plane scattering polarimeter was the balloon-borne mission \textsl{X-Calibur}~\citep{guo13,beilicke14,kislat18}, which will be succeeded by an improved version called \textsl{XL-Calibur} to be launched for the first time in 2022~\citep{abarr21}.
Other implementations of the principle include the Japanese small satellite \textsl{PolariS} \citep[Polarimetry Satellite,][]{hayashida14} and the Compton X-ray Polarimeter~\citep[\textsl{CXPOL},][]{chattopadhyay13,chattopadhyay14_2,chattopadhyay15} currently under consideration for the next Indian astronomy satellite.
All of these polarimeters consist of a compact low-$Z$ scattering element surrounded by high-$Z$ photon detectors, but details of the implementation vary.
The detector principle is made possible by the development of grazing incidence focusing optics for hard X-rays, such as the \textsl{InFOC$\mu$S}~\citep{tueller_etal_exa_2005}, \textsl{NuSTAR}~\citep{harrison05}, and \textsl{Astro-H}~\citep{kunieda10} telescopes.

\begin{figure}[t]
  \centering
  \includegraphics[width=\textwidth]{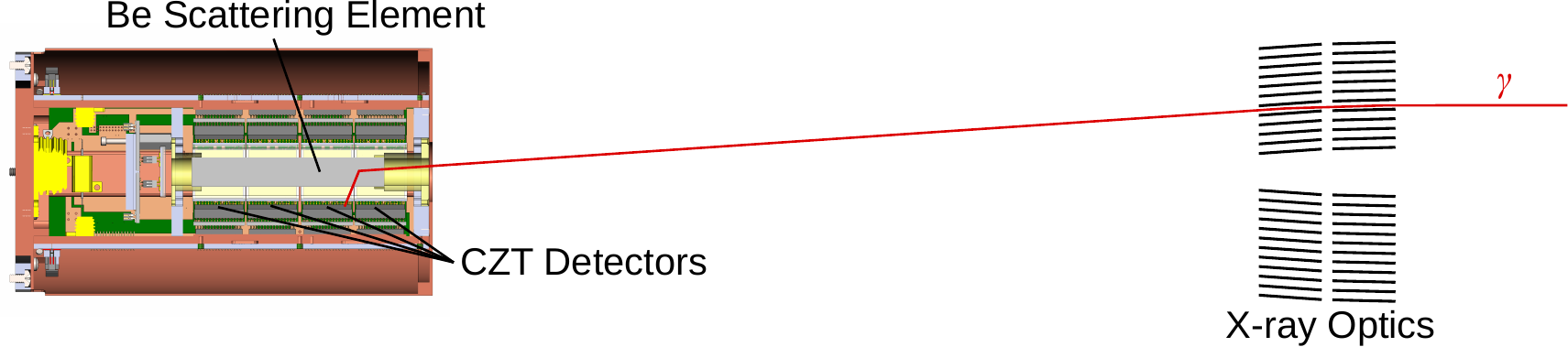}
  \caption{Focal plane scattering polarimeters consist of a compact scattering element surrounded by photon detectors at the focal spot of an X-ray mirror. The scattering element may either be passive, such as the Be scattering element of \textsl{XL-Calibur}, or it can be an active detector component, such as plastic scintillator. Various types of photon detectors surrounding the scattering element are being used, such as inorganic scintillators or CZT detectors as shown.}
  \label{fig:xl-calibur}
\end{figure}

\textsl{X-Calibur} and \textsl{XL-Calibur} use a passive beryllium scattering element surrounded by pixelated CZT detectors, whereas \textsl{PolariS} and \textsl{CXPOL} both use plastic scintillator to scatter photons surrounded by high-$Z$ inorganic scintillators as photon detectors.
In case of \textsl{PolariS}, the scattering element consists of an \num{8x8} array of \SI[product-units=power]{2.1x2.1x40}{mm} bars read out by a multi-anode PMT enabling imaging polarimetry.
Figure~\ref{fig:xl-calibur} shows a view of the \textsl{XL-Calibur} detector to illustrate the concept.

In each case, the choice of scattering and absorber material is a result of a trade-off between various design goals.
The active scattering element of \textsl{PolariS} and \textsl{CXPOL} is a powerful tool to reject background, due to its compact size.
For example, during its first engineering flight, \textsl{X-Calibur} had an active plastic scintillator scattering element.
Requiring a coincidence between signals in the scattering element and the surrounding CZT detectors reduced the residual background rate by a factor~20 in addition to the background suppression due to the CsI anti-coincidence shield surrounding the detector~\citep{beilicke15}.
In case of \textsl{CXPOL}, active coincidence is also necessary in order to reduce the background due to the high dark count rate of the SiPM readout \citep{chattopadhyay15}.
Additionally, in case of \textsl{PolariS}, the segmented scattering element enables a more precise reconstruction of the azimuthal scattering angle.
However, an active scattering element requires a rather large energy deposit.
For example, laboratory testing of a prototype for \textsl{CXPOL} found a \SI{\sim 6}{\percent} detection efficiency at \SI{0.5}{keV} energy deposit, with \SI{100}{\percent} efficiency reached at an energy deposition of about \SI{7}{keV}~\citep{chattopadhyay14_2}.
These energy depositions correspond to the energy loss of a \ang{90} scattered photon of \SI{16}{keV} and \SI{63}{keV}, respectively.

An optimization study for \textsl{X-Calibur} found that requiring a positive coincidence with the scattering element reduces polarization sensitivity due to the low trigger efficiency for signal events.
A passive Be scattering element was chosen as a replacement for the scintillator due to its lower average atomic number.
In the \SIrange{12}{27}{keV} energy range, the fraction of scattered photons was found to be \SI{45}{\percent} higher in Be compared to plastic scintillator~\citep{kislat18}.
Furthermore, this choice enabled a more compact detector design due to the \SI{80}{\percent} higher density of Be compared to plastic.
It is important to note that the results of this trade-off are specific to \textsl{X-Calibur} and \textsl{XL-Calibur}.
They depend on the effective area of the X-ray optics, the background environment, and the fact the instruments are balloon-borne, which influences the effective area due to absorption in the residual atmosphere above the payload.

Requirements for the photon-absorbing detectors surrounding the scattering element include the ability to measure the azimuthal scattering angle, detection efficiency, energy resolution, and readout complexity.
For example, the plastic scintillator element in \textsl{CXPOL} is surrounded by a circular arrangement of 32 CsI scintillator bars, providing optimized azimuthal symmetry.
On the other hand, \textsl{PolariS} and \textsl{XL-Calibur} rely on rotation of the instrument to average out the asymmetry of their square detector arrangements.
In contrast to the other two instruments, the pixelated CZT detectors of \textsl{XL-Calibur} not only allow a measurement of the azimuth angle, but also determine the position of the scattered photon along the optical axis.
The latter can serve as a proxy for the scattering angle, potentially resulting in an improved polarization sensitivity~\citep{krawczynski_2011}.
Furthermore, the ${\lesssim}\SI{10}{\percent}$ energy resolution of the CZT detectors is superior to the scintillator-based instruments.
On the other hand, \textsl{XL-Calibur} has a total of 1024 readout channels for the polarimeter, plus 64 channels of an imaging (but not polarization sensitive) CZT detector, resulting in a significantly higher complexity.

A general advantage of focal plane instruments over other polarimeter designs is their compact size.
This compact size intrinsically results in a reduced background rate.
Furthermore, it makes thick anticoincidence shields possible.
For example, \textsl{XL-Calibur} is completely surrounded by a \SIrange{3}{4}{cm} thick bismuth germanate (BGO) shield, except for a \SI{45}{mm}-diameter entrance hole~\citep{abarr21}. The total BGO mass is \SI{42}{kg}, and in case of larger detector designs a similarly thick shield would be prohibitively heavy.

However, combination of a scattering polarimeter with focusing X-ray optics significantly increases the system complexity.
The scattering elements of \textsl{XL-Calibur} and \textsl{CXPOL} have a diameter of \SI{12}{mm} and \SI{10}{mm}, respectively, and the plastic scintillator array of \textsl{PolariS} is quadratic with a side length of \SI{18}{mm}.
A synchrotron beam test of \textsl{X-Calibur} found that an alignment of the focal spot with the center of the scattering element within \SI{3}{mm} is required in order to keep the systematic uncertainty at \SI{<1}{\percent}~\citep{beilicke14}.
This alignment requirement places strict requirements on the design of the optical bench considering the focal length of typical X-ray optics (\textsl{X-Calibur}: \SI{8}{m}, \textsl{XL-Calibur}: \SI{12}{m}, \textsl{PolariS}: \SI{6}{m}, \textsl{CXPOL}: \SI{10}{m}).
Consideration must be given to both thermal deformations and at least in case of balloon-borne telescopes bending due to the changing influence of gravity as the telescope points at different elevations~\citep{kislat17}.
Along with the mechanical alignment of the optical bench comes a similarly strict absolute pointing, pointing stability, and pointing knowledge requirement.
For example, \textsl{XL-Calibur} requires a pointing knowledge ${<}\ang{;;15}$ corresponding to \SI{0.9}{mm} in the focal plane.

While plastic scintillator is the most common choice for scattering elements current focal plane polarimeter design, development of new detector technologies might enable completely new designs.
For example, recent improvements in the development of Silicon based fast active pixel sensors like DEPFETs~\citep{bombelli09_depfet}, Hybrid CMOS detectors like SPEEDSTERs~\citep{griffith16,chattopadhyay18_HCDoverview} with thicker depletion layer,
could allow `all-in-one' instruments enabling imaging hard X-ray and soft $\gamma$-ray polarimetry in the \SIrange{10}{80}{keV} energy range~\citep{muleri12,vadawale12}.

Development of focusing optics like the Laue lens \citep{barriere11} or newly proposed stacked prism lenses \citep{Mi19_spl} and channelling optics~\citep{shirazi_2018} to concentrate photons above \SI{100}{keV} have the potential to significantly increase the sensitivity of $\gamma$-ray polarimeters.

\section{Pair Production polarimetry}\label{sec:pair}
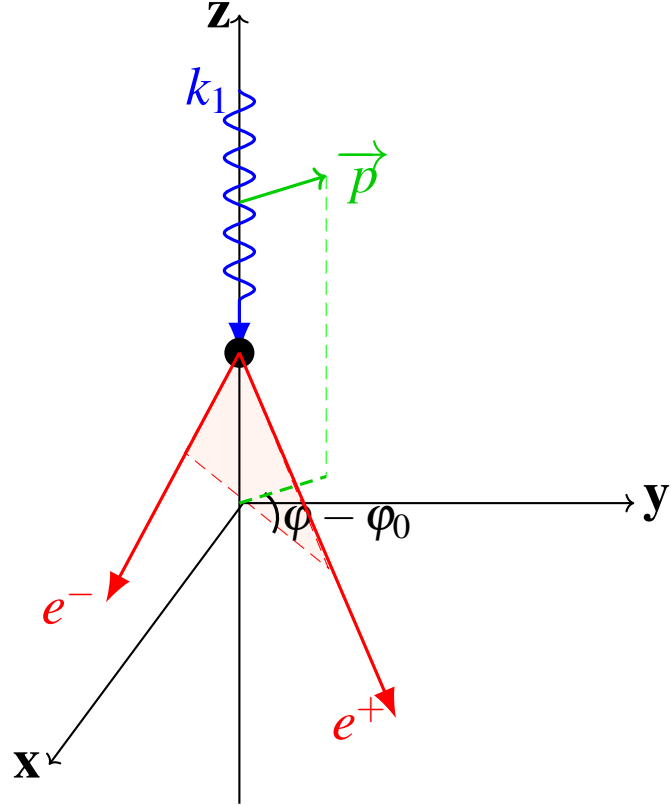
\begin{figure}
\centering
    
\begin{tikzpicture}
\pgfmathsetmacro{\cubex}{0.5}
\pgfmathsetmacro{\cubey}{6}
\pgfmathsetmacro{\cubez}{0.5}
\pgfmathsetmacro{\offsetx}{2}
\pgfmathsetmacro{\offsety}{-2}
\pgfmathsetmacro{\offsetz}{-2}

\definecolor{orange_shade}{rgb}{0.98,0.3,0.0}
\colorlet{orange_shade2}{orange_shade!6}

\draw [red,fill=orange_shade2, dash pattern=on5pt off3pt](-\cubex/2,-1,-\cubez/2) node[anchor=north]{}
  -- (-1.65,-3,-2.0) node[anchor=north]{}
  -- (1.8,-3.0,2.0) node[anchor=south]{}
  -- cycle;

\draw [black, thick, <-] (-\cubex/2,3.5,-\cubez/2) node[left]{{\huge\textbf z}} -- (-\cubex/2,-7,-\cubez/2);
\draw [black, thick, ->] (-0.2,-3,-\cubez/2) -- (5,-3,-\cubez/2) node[right]{{\huge\textbf y}};
\draw [black, thick, ->] (-0.2,-3,-\cubez/2)-- (-2.3,-6,1) node[left]{{\huge\textbf x}};
\coordinate (origin) at (-\cubex/2,-3,-\cubex/2);

\coordinate (scat1) at (-\cubex/2,-1,-\cubez/2);
\coordinate (scat2) at(\offsetx+5,\offsety-.5,\offsetz+7);

\draw [blue, very thick,-{Latex[length=4mm]},decorate, decoration={snake,amplitude=2mm,segment length=5mm,post length=5mm}, label=left:Incident photon] (-\cubex/2,2.5,-\cubez/2) node[left]{{\huge $\boldsymbol{k_1}$}} -- (scat1);

\fill[black] (scat1) circle (0.2);

\draw [black, very thick] (0.3,-3.3,0) arc (-30:55:0.4) ;
\draw (0.2,-3.25,-\cubex/2) node[black, right]{{\LARGE $\varphi - \varphi_0$}};

\draw [red, very thick,-{Latex[length=4mm]},decorate, label=left:Scattered photon] (-\cubex/2,-1,-\cubez/2) -- (-2.7,-5,-2.0) node[left]{{\huge $\boldsymbol{e^-}$}};

\draw [red, very thick,-{Latex[length=4mm]},decorate, label=left:Scattered photon] (-\cubex/2,-1,-\cubez/2) -- (2.7,-5.0,2.0) node[left]{{\huge $\boldsymbol{e^+}$}};

\draw [green!80!black, very thick,->] (-\cubex/2,1,-\cubez/2) -- (1-\cubex/2,1.2,-\cubez/2-0.4) node[right]{{\huge $\overrightarrow{\boldsymbol{p}}$}};

\draw [green, thin, dash pattern=on5pt off3pt] (1-\cubex/2,1.2,-\cubez/2-0.4) -- (1-\cubex/2,-2.8,-\cubez/2-0.4);
\draw [green!80!black, very thick, dash pattern=on5pt off3pt] (-\cubex/2,-3,-\cubez/2) -- (1-\cubex/2,-2.8,-\cubez/2-0.4);

\end{tikzpicture}

\caption{Schema of a photon propagating along $-z$ and undergoing a conversion to an $e^+e^-$ pair.
$\varphi$ is the azimuthal angle of the event and
$\varphi_0$ is the polarization  angle of the incoming radiation. }
\label{fig:pairproduction}
\end{figure}

At energies above a several MeV, photons primarily interact via pair production (see Fig.~\ref{fig:scattering-cross-section}).
The direction in which the electron and positron are emitted is polarization sensitive as shown in Fig.~\ref{fig:pairproduction} and pair conversion can be used to measure polarization.
The main challenge is posed by multiple scattering of the leptons in the target material, and careful optimization is necessary to achieve sufficient target mass while keeping the mean free path long enough to be able to measure the lepton trajectories.

Pair conversion polarimetry provides insights into the physics of high-energy $\gamma$-ray emission, where emission mechanisms are expected to differ from those at play at lower energies.
For sensitivity reasons, the focus must be on the brightest sources in the $\gamma$-ray sky:
\begin{itemize}
\item Young pulsars, for which the energy threshold corresponding to
 the on-set of curvature radiation can be tagged by the observation
 of the sudden change in (linear) polarization fraction and
 polarization direction, by contrast with that of synchrotron
 radiation at lower energies 
 \citep[Figs. 5 and 6 of][]{Harding:2017ypk}.

\item Blazars, for which leptonic and hadronic
 models predict the same degree of X-ray polarization but a higher
 maximum $\gamma$-ray polarization in hadronic models than leptonic
 ones \citep{zhang13}.
\end{itemize}

\subsection{Differential cross section}\label{sub:pair:xsection}
The non-polarized differential cross section (DCS) for electron-positron pair production was first calculated 
by \citet{Bethe-Heitler}, and 
the full linearly-polarized DCS by \citet{May1951}.
The final state is described by five variables that can be chosen \citep{Bethe-Heitler} to be the
 polar angles $\theta_+$ and $\theta_-$, and the azimuthal angles
 $\varphi_+$ and $\varphi_-$, of the electron ($-$) and of the positron (+),
 respectively, and the fraction $x_+$ of the energy of the incident
 photon carried away by the positron, $x_+ \equiv E_+/E$.
The targets on which the photon converts are the charged particles
that are available in the detector:
\begin{itemize}
 \item $\gamma Z \to e^+e^-Z$ ~ (``nuclear'' conversion); 
 \item $\gamma e^- \to e^+e^-e^-$ ~  (``triplet'' conversion).
\end{itemize}
The DCS is 
independent on the nature (nucleus, electron) and on the mass of the target
(except for the screening of the electric field of the electron cloud,
described as a form-factor $F(q^2)$, $q$ momentum transferred to the
recoiling target).

\subsection{Polarization asymmetry}
Polarimetry is performed by the
analysis of the distribution of an azimuthal angle $\varphi$ that
measures the orientation of the final state in the azimuthal plane,
that is, in the plane orthogonal to the photon direction.
\begin{equation}
\gfrac{\dd N}{\dd \varphi} \propto 
\left(
1 + A \, P \cos[2(\varphi - \varphi_0)]
\right).
\label{eq:modulation}
\end{equation}
The polarization asymmetry, $A$, is a function of the photon energy, $E$ 
\citep{Yadigaroglu:1996qq}.
The angle $\varphi_0$ describes the polarization angle of the incoming
radiation.

As the conversion to a pair produces a three-particle final state, even in
the case when the recoil energy is so small that the track of the
recoiling particle cannot be measured, there is no natural definition
of ``the'' azimuthal angle.
In the case the direction of the incoming photon is not precisely
known, a common practice is to use $\omega$, the azimuthal angle of
the line that connects the electron to the positron in the azimuthal plane
\citep[Fig. 1 right of][]{Wojtsekhowski}.
In practice though, the direction of the incident photon is perfectly
known, for a point-like source, as $\gamma$-ray astronomy is performed
before $\gamma$-ray polarimetry is attempted, so the photon-to-source
assignment is available.
A better choice, that provides a significantly higher value of $A(E)$,
was found later to be the bisectrix of the azimuthal angles of the
electron and of the positron,
$\varphi_{+-} \equiv (\varphi_{+} + \varphi_{-})/2$ \cite{Gros:2016dmp}.
Note a $\pi/2$ phase difference between an $\omega$ analysis and the
$\varphi_{+-}$ analysis of the same data.
Incidently, the low-energy \citep[$A = \pi/4$,][]{Gros:2016dmp} 
and the high-energy \citep{Boldyshev:1972va}
asymptotic expressions for $A(E)$ were obtained from 
expressions of the integrated DCS that turned
out to be functions of $\varphi_{+-}$.

Above \SI{\sim 20}{\mega\electronvolt}, that is, over most of the
energy range accessible to experimentalists, the high-energy
asymptotic expression is found to be fairly accurate
\citep[Fig. 3 (a) of][]{Gros:2016dmp}.
At the highest energies, $A \approx 1/7$, as was already obtained by
\citet{Wick1951}.
The measurement of the polarization fraction $P$ and
direction $\varphi_0$ of the incoming radiation can be either
performed by a fit of the distribution of the azimuthal angle 
(eq. (\ref{eq:modulation})) or by using the moments' method
(Sect. 4 of \citet{Bernard:2013jea} and references therein,
 Sect. 4 of \citet{Gros:2016dmp}).

As the polarization asymmetry is found to be a strongly varying
variable on a number of kinetic variables that describe the final
state, e.g.,
on $x_+$, the fraction of the photon energy that is carried away by the positron
\citep[Fig. 1 left of][]{Wojtsekhowski};
on $\mu$, the pair invariant mass \citep[eq. (5) and Figs. 2-3 of][]{Bernard:2021wws};
on $q$ and on the pair opening angle $\theta_{+-}$ \citep[page 30 of][]{Integral2019},
it can be expected that using all the information present in the
five-dimensional final state provides a better sensitivity to polarization;
indeed, an improvement of a factor larger than two is at hand using the moments method on the full DCS
(Fig. 21 of \citet{Bernard:2013jea};
 Fig. 3 (a) of \citet{Gros:2016dmp}).
Due to the above-mentioned variation, any discrepancy between the real
and simulated acceptance of the detector, as a function of these
kinetic variables, would induce a systematic bias in the measurement,
and should be monitored.

\subsection{Multiple scattering}
A number of experimental effects affect the measurement of the
azimuthal angle, among which multiple scattering is certainly the
fiercest and has been one of the major obstacles to a polarimetry with
the past and present pair-conversion $\gamma$-ray telescopes.
The presence of a non-zero resolution $\sigma_\varphi$ on the azimuthal angle changes eq. 
 (\ref{eq:modulation}) to 
\begin{equation}
\gfrac{\dd N}{\dd \varphi} \propto 
\left(
1 + A \, e^{-2 \sigma_\varphi^2} P \cos[2(\varphi - \varphi_0)]
\right),
\label{eq:modulation:MS}
\end{equation}
so the effective polarization asymmetry is afflicted with a dilution factor 
$D = A_{\text{eff}} / A = e^{-2 \sigma_\varphi^2} $. 
The classical calculation \citep{Kelner} of
$\sigma_\varphi$ is performed under an approximation of the pair opening angle by 
the most probable value of its distribution~\citep{Olsen1963}.

Assuming an equal energy share, $p \approx E / (2c)$, 
the angular resolution for the azimuthal angle is then
$\sigma_\varphi = \sigma_0 \sqrt{x/X_0}$ with
$ \sigma_0 \approx \SI{24}{\radian} $~\citep[eq. (15) of][]{Bernard:2013jea},
where
$x$ and $X_0$ are the
thickness and the radiation length of the 
material traversed by the electron, respectively.
The dilution factor therefore reaches a value of  
$D = 1/2$ for a path length of $\approx \SI{110}{\micro\meter}$ in silicon.

The catastrophic exponential dependence of the dilution as a function
of thickness so obtained,
$D \approx \exp(- \sigma_0^2 x / (2 X_0))$
is not confirmed by the results of full simulations, actually,
(Fig. 17 of \citet{Bernard:2013jea}, Fig. 3 of \citet{Eingorn:2015oga}).
The dilution is found to be much larger (i.e. less degraded) at high thicknesses, 
than in the classical calculation, something that is the consequence of the $\theta_{+-}$
distribution having a huge tail at large $\theta_{+-}$ values \citep{Olsen1963}.

 The single-track angular resolution is not the simple contribution of multiple scattering in the conversion wafer.
An optimal fit \citep[such as a Kalman filter,][]{Fruhwirth:1987fm}
is performed that takes into account multiple
scattering and the space resolution of each of the detector layers
$\sigma$,
and the single-track angular resolution at the vertex is found to be
 $ \sigma_{\theta} \approx 
 \left(2 \sigma/d\right)^{1/4} 
 \left(e/X_0\right)^{3/8} 
 \left(p/p_0\right)^{-3/4}$,
$p_0$ is the multiple-scattering constant \citep{Zyla:2020zbs},
$d$  the longitudinal sampling (along the track),
$e$ the wafer thickness,
$X_0$ the material radiation length,
$\sigma$ the single-measurement track-position resolution.
Given the $1/E$ scaling of the $\theta_{+-}$ distribution
\citep{Olsen1963}, the induced dilution is then higher (less degraded)
at low energies~\citep[Fig. 20 of][]{Bernard:2013jea}.

\subsection{Polarimetry with Triplet Conversions}
The issues with the deleterious effects of multiple scattering on the
electron and positron of the pair have generated a number of studies
of triplet conversions, in the hope that the recoil of the 
electron at a large polar angle would help~\citep[][and references therein]{Boldyshev:1995ui}.
Unfortunately, most of triplet conversion produces recoiling electrons
with a very small momentum~\citep[Fig. 6 of][]{Bernard:2013jea} so the
performance of a polarimeter based on triplet conversion can be
expected to be quite poor~\citep[Sect. 5.3 of][]{Bernard:2013jea}.

\begin{figure}[t]
\hfill
\includegraphics[scale=.25]{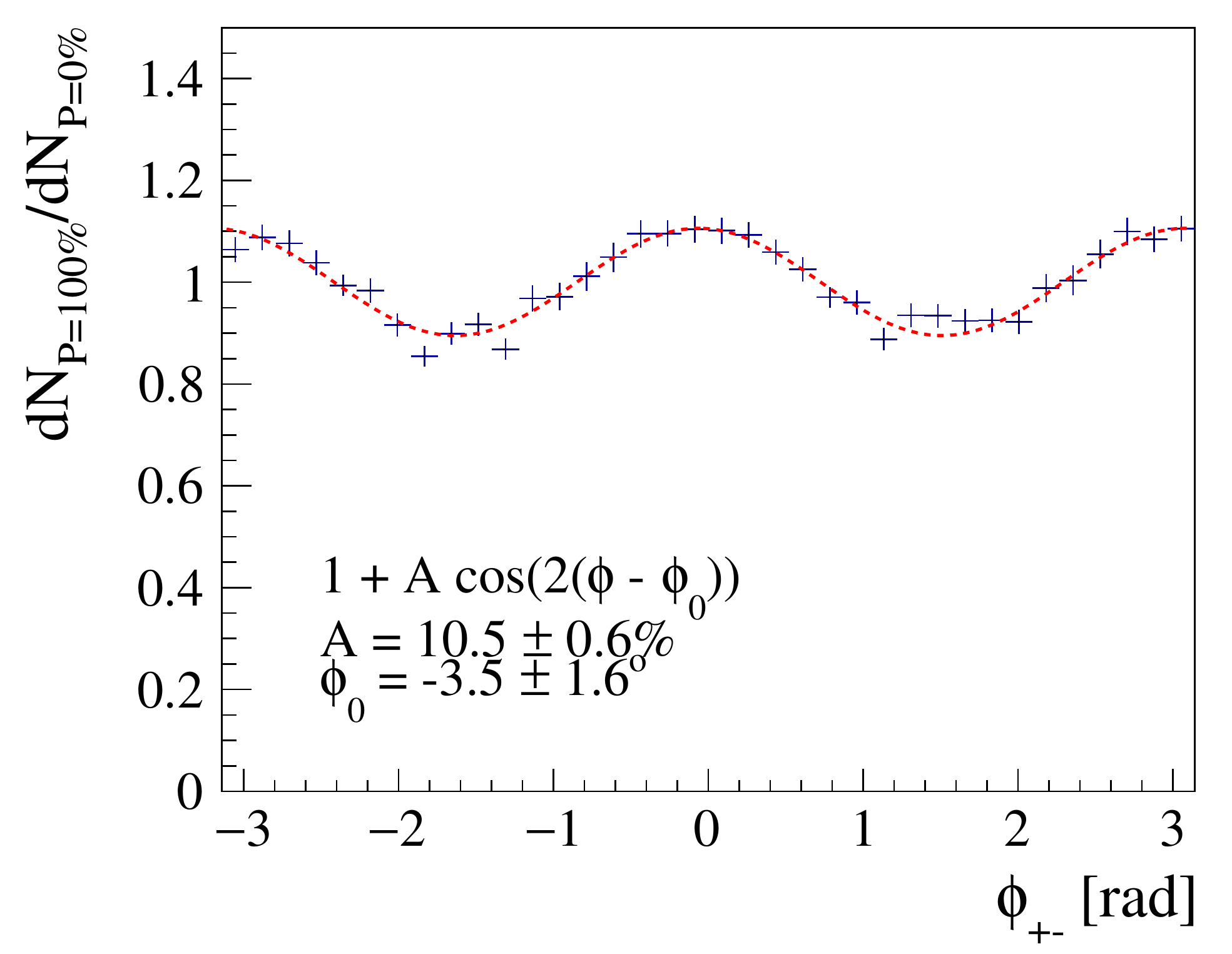}
\hfill
\includegraphics[scale=.25]{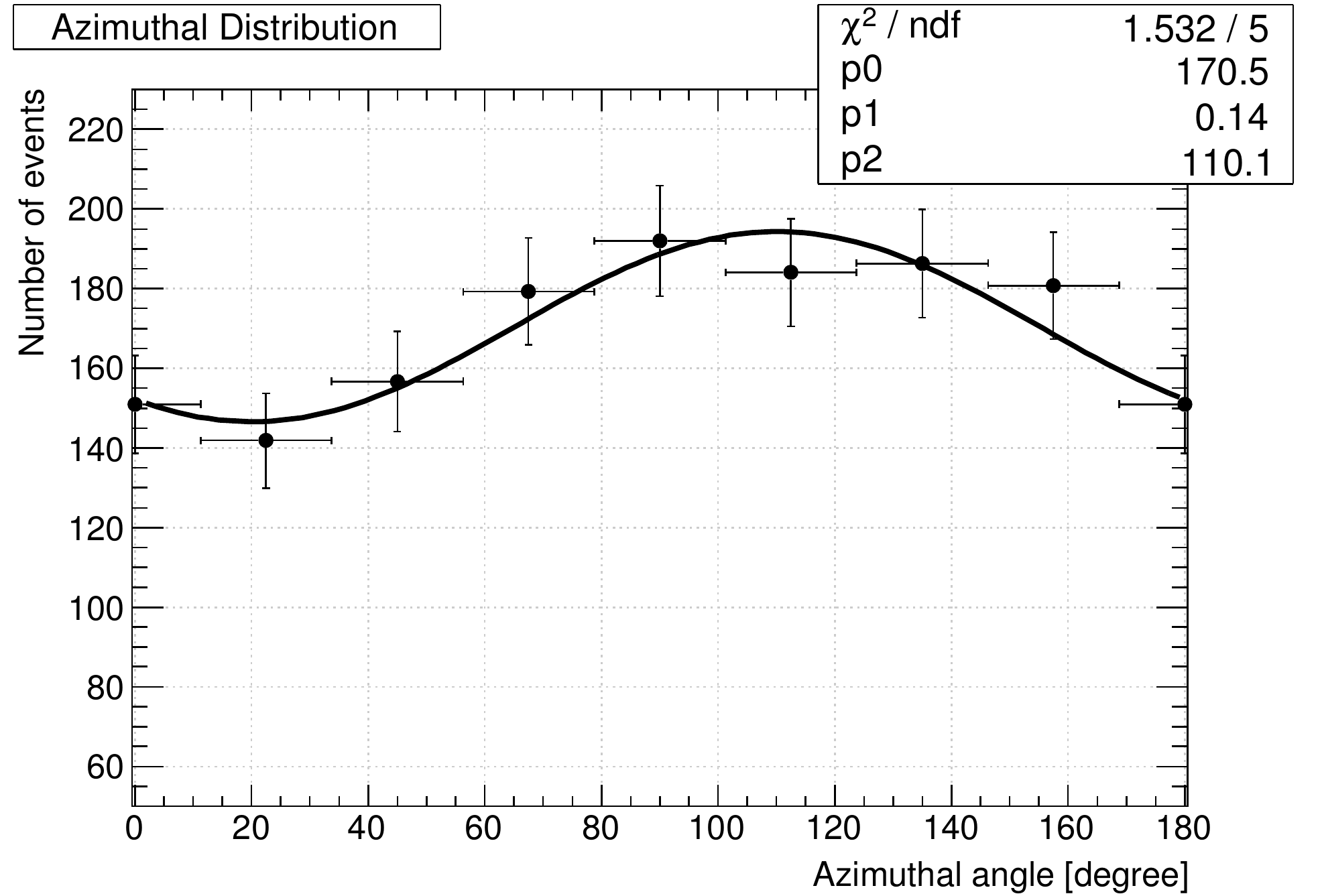}
\hfill ~
\caption{Pair-conversion $\gamma$-ray polarimetry:
  distribution of the azimuthal angle measured with polarimeter prototypes on fully linearly polarized beams.
  Left, HARPO time-projection chamber (TPC),  
  $E = \SI{12}{\mega\electronvolt}$ \citep{Gros:2017wyj}.
Right, GRAINE emulsions 
  $E = \SIrange{1.5}{2.4}{\giga\electronvolt}$
  \citep{Ozaki:2016gvw}.
Both with permission.
\label{pair:prototype:tests}}
\end{figure}

\subsection{Past experimental achievements }
Gamma-ray beams are currently produced in the laboratory using the
Compton scattering of a laser beam on a high-energy mono-energetic
electron beam.
Gamma-ray pseudo-monochromaticity is achieved by selecting the
``Compton-edge'' with a cylindrical collimator on the forward axis.
The linear polarization of the incident laser beam is then transferred
almost exactly to the $\gamma$-ray beam \citep{Sun:2011es}.

\paragraph{JLab prototype.}
The first experimental measurement  was the
characterization of a polarimeter prototype from a high-intensity
$\gamma$-ray beam for the Jefferson Laboratory.
\SIrange{1.5}{2.4}{GeV} polarized $\gamma$-rays from the LEPS beam line at
SPRing8 were converted to $e^+e^-$ in a \SI{0.1}{mm} thick Carbon foil with
\SI{0.02}{\percent} efficiency and a small angle acceptance, a configuration well
suited for the measurement on a small-emittance\footnote{In accelerator physics,
the emittance of a particle beam is a measure of its phase space volume in position and angle.}
$\gamma$-ray beam
but quite different from the specifications for a wide-field
detector.
The position of the electron and of the positron tracks were measured
with silicon-strip detectors (SSD) located several meters downstream
the converter  \citep{deJager:2007nf}.

\begin{figure}
\includegraphics[width=.4\linewidth, trim=4 110 90 80, clip]{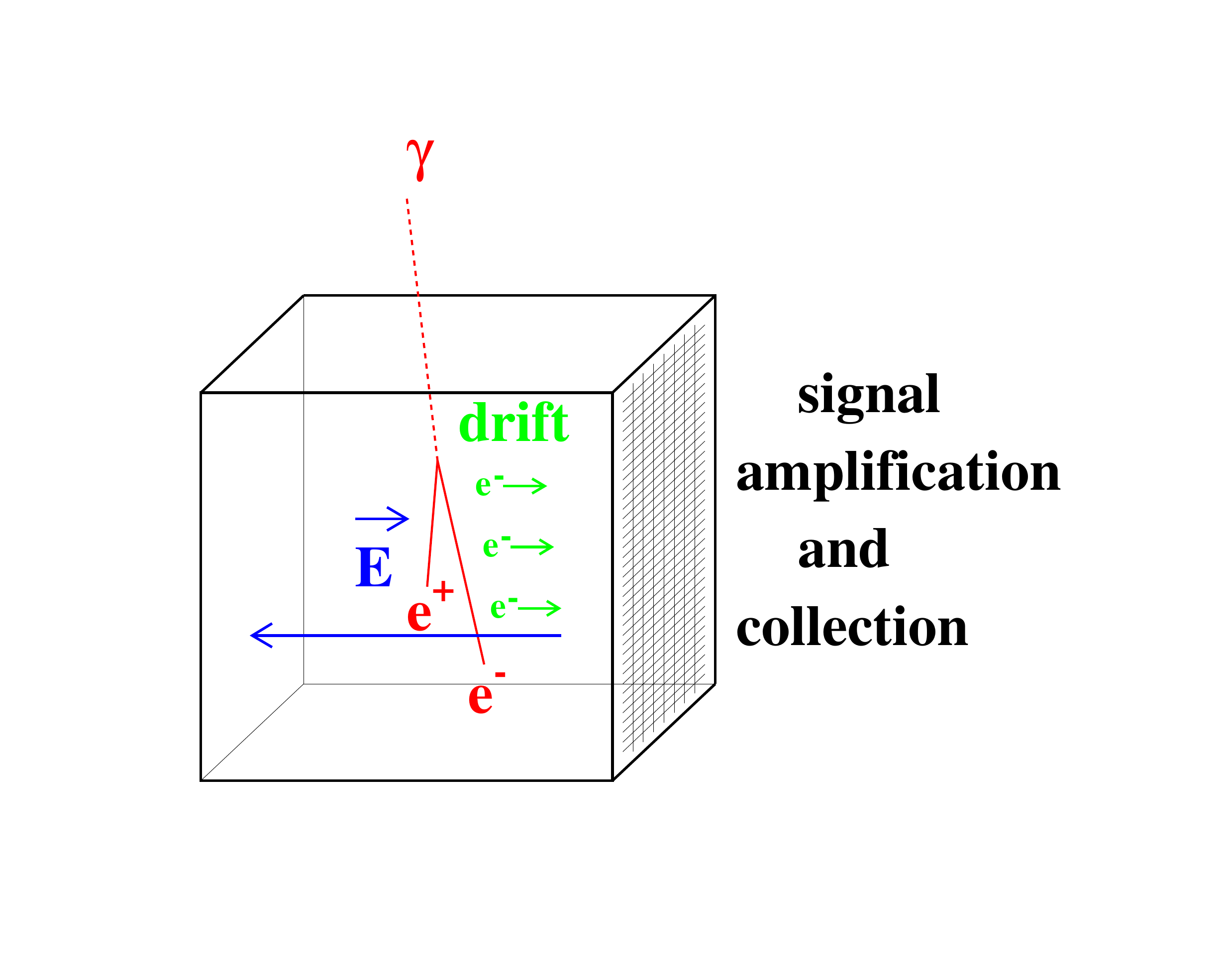} 
\hfill 
\includegraphics[width=.57\linewidth]{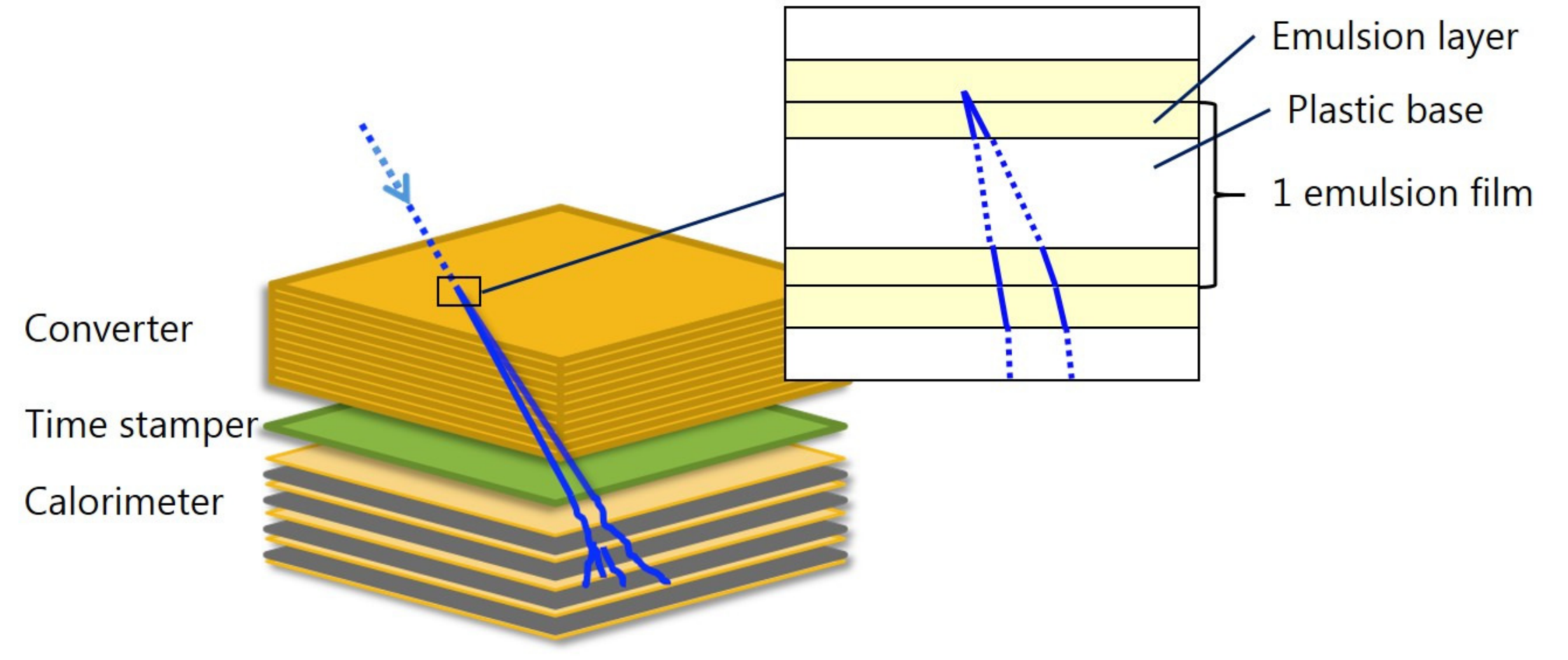} 
\caption{Two pair-polarimeters schemes:
  \emph{Left:} Time projection chamber (TPC).
  \emph{Right:} Emulsion detector \citep[with permission,][]{Ozaki:2016gvw}.
}
\label{fig:pair:polarimeters}
\end{figure}

\paragraph{GRAINE.}
The GRAINE project (Gamma-Ray Astro-Imager with Nuclear Emulsion)
develops an emulsion-based active target to perform $\gamma$-ray
astronomy and polarimetry in the \SI{10}{MeV} -- \SI{100}{GeV} energy range
\citep{Takahashi:2015jza}, Fig. \ref{fig:pair:polarimeters}, right.
An emulsion is a high-density, sub-micron resolution homogeneous detector.
A GRAINE prototype has been characterized on the LEPS beam
\cite[Fig. \ref{pair:prototype:tests} right and][]{Ozaki:2016gvw}.

\paragraph{HARPO.}
The HARPO project (Hermetic ARgon POlarimeter) has developed a gas TPC
prototype 
\citep{Bernard:2018jql}, Fig. \ref{fig:pair:polarimeters} left,
and characterized it on the 1.7 to 74 MeV $\gamma$-ray beam of the BL01 beam line at LASTI
\citep[Fig. \ref{pair:prototype:tests}, left and][]{Gros:2017wyj}.

~

The three projects have been able to measure the polarization of the
beam with an excellent value of the dilution factor.

\subsection{Future prospects}
\paragraph{Gas TPCs.} 

The possible use of gas TPCs for $\gamma$-ray polarimetry is detailed
in Chapter ``Time projection Chambers for gamma-ray astronomy''. These
devices provide an excellent imaging capability down to low energies,
close to the conversion threshold.
The degradation of the gas quality with time, that proved to be an
issue for EGRET spark chambers~\citep{EGRET:1999}, should not be a problem because
proportional gas counters show a much lower polymerization rate
than spark chambers.
The design of ageing-resistant detectors benefits from the body
of expertise that has been developed for the high-energy physics at
the large hadron collider (LHC), with dose rates a factor of thousands
to millions times larger than for detectors on a space mission~\citep[see the review by][]{Titov:2004rt}.
In most high-energy-physics (HEP) experiments that are using a TPC, the start time of the electron drift is determined from a trigger that is generated using the information provided by other sub-detectors of the experiment. 
For a large (multi-meter-cube) TPC used for astrophysical $\gamma$-ray detection this might not be a option due to the limitation of the mass budget on a space mission, and a self-trigger operation would be required.
In a telescope consisting of a multi-module active target \citep{Bernard:2014kwa}, a trigger could be formed from the 
simultaneous crossing of the anodes of several modules, with a precision of a couple of nanoseconds.
A fast pseudo-tracking can also be used to build a level-two trigger signal 
using the real time information
provided by a dedicated electronics \citep{AGET}.
An alternative, for a stand-alone TPC, is a trigger-less, continuous
mode \citep{Rauch:2012hp,Hunter:2013wla}.

\paragraph{Liquid or Solid TPCs.} 

Given the volume limitation on a space mission using a liquid detector
seems to be tempting \citep{Caliandro:2013kba}, with a density gain of
a factor of 840 for Argon with respect to $\SI{1}{\bar}$ gas.
However, the performance of the tracking system must scale accordingly, which presents several difficulties.
In particular, the typical collected electrical charge is left
unchanged by the double scaling (same collected charge on a \SI{1}{cm}
track segment in \SI{1}{\bar} gas as on a \SI{12}{\micro\meter} segment in
a liquid).
An issue arises from the fact that charge  amplification can take place in a gas but not in a liquid.
Therefore single-phase liquid-argon detectors do not allow such a small scale tracking.
On Earth, the problem can be overcome by the use of a two-phase system. In that approach, the active target consists of liquid argon, from which the electrons are extracted for amplification in the gas phase.
In zero gravity on orbit, the stability of the gas-liquid
interface might be an issue though, that could be mitigated by the use of a
solid TPC, with similar electron transport properties as in a liquid~\citep{Aprile:1985xz}.
Another major issue is the diffusion of the electrons during drift, 
as the diffusion coefficient saturates in
liquid argon for high electric field, at values
$\SI{\sim 100}{\micro\meter}/\sqrt{\si{\centi\meter}}$, 
that is, similar to that in gas (with a quencher).
The diffusion too, therefore, does not scale with density.
Overall, it seems unlikely that the imaging of photon conversions 
can be performed with liquid or solid TPCs,
at the ${<} 10^{-3} X_0$ scale that is needed for polarimetry.

\paragraph{Emulsions}

Photographic emulsions enable a sub-micron imaging of photon
conversions in a dense (solid) active targets as demonstrated by
GRAINE, and current balloons are able to fly multi-ton telescopes at
operation altitude.
The technique therefore shows an impressive potential for $\gamma$-ray
polarimetry.
The main limitations are the difficulty to trigger / select /
reconstruct low-energy conversions (the range is currently limited to
\SIrange{>50}{100}{\mega\electronvolt}), the difficulty to analyze short
transients, with a timestamper that provides a second-scale timing
precision, and the need to expose the emulsions to the ambient cosmic
rays for the shortest time possible \citep{Takahashi:2015jza}.

\paragraph{Silicon detectors}

Present and planned $\gamma$-ray telescopes are using an active target
that consists of a stack of SSDs, in some cases interleaved with
high-$Z$ converters such as tungsten foils to improve the effective
area.
No significant demonstration that a non-zero  effective polarization asymmetry can be achieved with SSDs has been published to date, be it  by the analysis of simulated data or of prototype tests on beam.
This is why, most likely, the polarimetry estimates that have been published are based on
assumed values of the effective polarization asymmetry of the
detectors \citep{DeAngelis:2016slk,Giomi:2016brf}.

\subsection{Effective area and sensitivity}

In contrast to ``thick'' active targets, in which the conversion
probability of the photon is close to unity and the effective area is
the product of the geometric area by the efficiency, for ``thin''
active targets, it is better expressed as the product of the detector
sensitive mass, $M$, by the attenuation coefficient, $H$, which is
found to be approximately proportional to $Z^2/A$ \citep[atomic number and
mass, respectively,][]{NIST:gamma}.
For a perfect detector with 100\,\% efficiency, 
and for a  ``Crab-like'' source with a spectral index of $\Gamma = 2$,
with flux $f(E) = F_0 / E^2$ and
$F_0 = \SI{1e-3}{\mega\electronvolt\per\square\centi\meter\per\second}$,
the number of events is 
 $N =  T M F_0 \int {H(E)} \ {E^2} \dd E $, 
where $T$ is the mission duration.
The variation of $f(E) H(E)$ as a function of $E$ is available in Fig. 
2 of \citet{Bernard:2013jea}.
The integral has a mild variation with the nature of the active target, 
varying from 
 1.9 \si{\centi\meter^2 / (\kilo\gram \, \mega \electronvolt)} for neon to
 8.3 \si{\centi\meter^2 / (\kilo\gram \, \mega \electronvolt)} for xenon.
A $\SI{1}{\kilogram}\cdot\annee$ argon mission with full efficiency,
acceptance, exposure and perfect dilution down to $2 m c^2$,
 would observe $N \approx 10^5$ events with an average polarization asymmetry of
$\langle A \rangle\approx 0.33$ and, therefore, a precision of the measurement of $P$,  of $\sigma_P \approx \sqrt{2/N} / \langle A \rangle \approx  0.0135 $. 

More realistically, a $\SI{1}{\kilogram}\cdot\annee$ argon-mission with 
a \SI{10}{\mega\electronvolt} threshold,
a \SI{10}{\percent}-exposure-and-efficiency above threshold and
a $D=0.5$ dilution, 
would detect $N\approx 5000$ events with 
$\langle A \rangle\approx 0.232$,
$\langle A_\text{eff} \rangle\approx 0.116$,
and
$\sigma_P \approx  0.17$. 
Therefore the polarimetry of a cosmic source with pairs should focus
on the brightest sources of the MeV sky, in the first place, and with
a good-dilution, low-threshold, large-sensitive mass detector in orbit
for several years.

\section{Summary and Outlook}\label{sec:summary}
The gamma-ray sky shows a wide variety of  sources undergoing
amongst the wildest phenomena in the Universe.
These gamma-rays cannot be produced by thermal emission, instead most
often by processes that either produce linearly polarized radiation,
such as synchrotron radiation or curvature radiation, or that conserve
(some amount of) polarization, such as (inverse) Compton scattering.
In that polarized context, the hadronic interactions of cosmic rays on
matter at rest can be singled out by their producing neutral pions
that decay to non polarized photons.
Gamma-ray polarimetry would therefore be an extremely powerful
diagnostic to complement our understanding of the physical mechanisms
in these sources to date.

Unfortunately, forty years after the precise, significant measurement
of the OSO-8 measurement on the Crab nebula in the X-ray band
\citep{weisskopf78}, and despite the abounding creativity and the
immense efforts deployed by the community, there is no equivalent
result in the gamma-ray band.

This chapter reviews the achievements, the available measurement technologies and the future prospects in the field. With the upcoming launch of the \textsl{POLAR-2} detector, with a sensitivity
increased by one order of magnitude compared to \textsl{POLAR}, the recent selection of
COSI as a Small Explorer mission by NASA, and the development of the \textsl{ASTROMEV} (the successor of \textsl{e-ASTROGAM}) and \textsl{AMEGO} projects, there is good hope that we can see some polarized
light at the end of the tunnel.


%
%
%

\section{Cross-References}
\begin{itemize}
\item Volume 1 X-rays
  
  \begin{itemize}
  \item Section Detectors for X-ray Astrophysics
    \begin{itemize}
    \item Time projection for polarimetry
    \item Compton polarimetry
    \item SDD and silicon strip
    \item High-Z semi-conductors (CdZn, CdZnTe)
    \end{itemize}
  \item Section X-ray Missions
    \begin{itemize}
    \item Astrosat
    \end{itemize}
  \end{itemize}
  
\item Volume 2 Gamma-rays
  
  \begin{itemize}
  \item Section Optics and Detectors for Gamma-ray Astrophysics
    \begin{itemize}
    \item Telescope concepts in gamma-ray astronomy
    \item Coded mask instruments
    \item Compton telescopes
    \item Germanium detectors for gamma-ray astronomy
    \item Silicon detectors for gamma-ray astronomy
    \item CdTe and CZT detectors for gamma-ray astronomy
    \item Scintillators for gamma-ray astronomy
    \item Photodetectors for gamma-ray astronomy
    \item Time projection chambers for gamma-ray astronomy
    \item Detector and mission design simulations
    \end{itemize}

  \item Section Space-based Gamma-ray Observatories
    \begin{itemize}
    \item The COMPTEL instrument on the CGRO mission
    \item
      The INTErnational Gamma-Ray Astrophysics Laboratory (INTEGRAL)
    \end{itemize}

  \item Section Polarimetry, in fact devoted to Analyis.
    \begin{itemize}
    \item Bayesian analysis of the data from PoGo+
    \item Analysis of the data from POLAR
    \item Soft gamma-ray polarimetry with COSI using maximum likelihood analysis
    \item Stokes parameter analysis of XL-Calibur data
    \end{itemize}
  \end{itemize}
\end{itemize}

\bibliographystyle{spbasic}
\bibliography{reference}

\end{document}